\documentclass[a4paper,11pt]{article}
\pdfoutput=1 

\usepackage{lineno}
\usepackage{jinstpub} 
\usepackage{xspace}
\usepackage{makecell}
\usepackage{verbatim}

\usepackage{pbox}

\title{Measurement of scintillation response of CsI[Na] to low-energy nuclear recoils by COHERENT}

\newcommand{\Mephi}{a}
\newcommand{\Mephidesc}{\affiliation[\Mephi]{National Research Nuclear University MEPhI (Moscow Engineering Physics Institute), Moscow, 115409, Russian Federation}}
\newcommand{\Duke}{b}
\newcommand{\Dukedesc}{\affiliation[\Duke]{Department of Physics, Duke University, Durham, NC, 27708, USA}}
\newcommand{\TUNL}{c}
\newcommand{\TUNLdesc}{\affiliation[\TUNL]{Triangle Universities Nuclear Laboratory, Durham, NC, 27708, USA}}
\newcommand{\UTK}{d}
\newcommand{\UTKdesc}{\affiliation[\UTK]{Department of Physics and Astronomy, University of Tennessee, Knoxville, TN, 37996, USA}}
\newcommand{\ITEP}{e}
\newcommand{\ITEPdesc}{\affiliation[\ITEP]{Institute for Theoretical and Experimental Physics named by A.I. Alikhanov of National Research Centre ``Kurchatov Institute'', Moscow, 117218, Russian Federation}}
\newcommand{\ORNL}{f}
\newcommand{\ORNLdesc}{\affiliation[\ORNL]{Oak Ridge National Laboratory, Oak Ridge, TN, 37831, USA}}
\newcommand{\USD}{g}
\newcommand{\USDdesc}{\affiliation[\USD]{Physics Department, University of South Dakota, Vermillion, SD, 57069, USA}}
\newcommand{\NCSU}{h}
\newcommand{\NCSUdesc}{\affiliation[\NCSU]{Department of Physics, North Carolina State University, Raleigh, NC, 27695, USA}}
\newcommand{\Sandia}{i}
\newcommand{\Sandiadesc}{\affiliation[\Sandia]{Sandia National Laboratories, Livermore, CA, 94550, USA}}
\newcommand{\UW}{j}
\newcommand{\UWdesc}{\affiliation[\UW]{Center for Experimental Nuclear Physics and Astrophysics \& Department of Physics, University of Washington, Seattle, WA, 98195, USA}}
\newcommand{\LANL}{k}
\newcommand{\LANLdesc}{\affiliation[\LANL]{Los Alamos National Laboratory, Los Alamos, NM, 87545, USA}}
\newcommand{\Laurentian}{l}
\newcommand{\Laurentiandesc}{\affiliation[\Laurentian]{Department of Physics, Laurentian University, Sudbury, Ontario, P3E 2C6, Canada}}
\newcommand{\CMU}{m}
\newcommand{\CMUdesc}{\affiliation[\CMU]{Department of Physics, Carnegie Mellon University, Pittsburgh, PA, 15213, USA}}
\newcommand{\IU}{n}
\newcommand{\IUdesc}{\affiliation[\IU]{Department of Physics, Indiana University, Bloomington, IN, 47405, USA}}
\newcommand{\VT}{o}
\newcommand{\VTdesc}{\affiliation[\VT]{Center for Neutrino Physics, Virginia Tech, Blacksburg, VA, 24061, USA}}
\newcommand{\NCCU}{p}
\newcommand{\NCCUdesc}{\affiliation[\NCCU]{Department of Mathematics and Physics, North Carolina Central University, Durham, NC, 27707, USA}}
\newcommand{\INR}{q}
\newcommand{\INRdesc}{\affiliation[\INR]{Institute for Nuclear Research of the Russian Academy of Sciences, Moscow, 117312, Russian Federation}}
\newcommand{\UF}{r}
\newcommand{\UFdesc}{\affiliation[\UF]{Department of Physics, University of Florida, Gainesville, FL, 32611, USA}}
\newcommand{\Tufts}{s}
\newcommand{\Tuftsdesc}{\affiliation[\Tufts]{Department of Physics and Astronomy, Tufts University, Medford, MA, 02155, USA}}
\newcommand{\SNU}{t}
\newcommand{\SNUdesc}{\affiliation[\SNU]{Department of Physics and Astronomy, Seoul National University, Seoul, 08826, Korea}}
\author[\Mephi]{D.~Akimov,}\Mephidesc
\author[\Duke,\TUNL]{P.~An,}\Dukedesc\TUNLdesc
\author[\Duke,\TUNL]{C.~Awe,}
\author[\Duke,\TUNL]{P.S.~Barbeau,}
\author[\UTK]{B.~Becker,}\UTKdesc
\author[\ITEP,\Mephi]{V.~Belov ,}\ITEPdesc
\author[\UTK]{I.~Bernardi,}
\author[\ORNL]{M.A.~Blackston,}\ORNLdesc
\author[\USD]{C.~Bock,}\USDdesc
\author[\Mephi]{A.~Bolozdynya,}
\author[\NCSU]{J.~Browning,}\NCSUdesc
\author[\Sandia]{B.~Cabrera-Palmer,}\Sandiadesc
\author[\USD,1]{D.~Chernyak,}\note{Now at: Department of Physics and Astronomy, University of Alabama, Tuscaloosa, AL, 35487, USA and Institute for Nuclear Research of NASU, Kyiv, 03028, Ukraine}
\author[\Duke]{E.~Conley,}
\author[\ORNL]{J.~Daughhetee,}
\author[\UW]{J.~Detwiler,}\UWdesc
\author[\USD]{K.~Ding,}
\author[\UW]{M.R.~Durand,}
\author[\UTK,\ORNL]{Y.~Efremenko,}
\author[\LANL]{S.R.~Elliott,}\LANLdesc
\author[\ORNL]{L.~Fabris,}
\author[\ORNL]{M.~Febbraro,}
\author[\Laurentian]{A.~Gallo Rosso,}\Laurentiandesc
\author[\ORNL,\UTK]{A.~Galindo-Uribarri,}
\author[\TUNL,\ORNL,\NCSU]{M.P.~Green ,}
\author[\ORNL]{M.R.~Heath,}
\author[\Duke,\TUNL]{S.~Hedges,}
\author[\CMU]{D.~Hoang,}\CMUdesc
\author[\IU]{M.~Hughes,}\IUdesc
\author[\Duke,\TUNL]{T.~Johnson,}
\author[\Mephi]{A.~Khromov,}
\author[\Mephi,\ITEP]{A.~Konovalov,}
\author[\Mephi,\ITEP]{E.~Kozlova,}
\author[\Mephi]{A.~Kumpan,}
\author[\Duke,\TUNL]{L.~Li,}
\author[\VT]{J.M.~Link,}\VTdesc
\author[\USD]{J.~Liu,}
\author[\NCSU]{K.~Mann,}
\author[\NCCU,\TUNL]{D.M.~Markoff,}\NCCUdesc
\author[\IU]{J.~Mastroberti,}
\author[\INR]{Y.A.~Melikyan,}\INRdesc
\author[\ORNL]{P.E.~Mueller,}
\author[\ORNL]{J.~Newby,}
\author[\CMU]{D.S.~Parno,}
\author[\ORNL]{S.I.~Penttila,}
\author[\Duke]{D.~Pershey,}
\author[\CMU]{R.~Rapp,}
\author[\UF]{H.~Ray,}\UFdesc
\author[\Duke]{J.~Raybern,}
\author[\Mephi,\ITEP]{O.~Razuvaeva,}
\author[\Sandia]{D.~Reyna,}
\author[\TUNL]{G.C.~Rich,}
\author[\NCCU,\TUNL]{J.~Ross,}
\author[\Mephi]{D.~Rudik,}
\author[\Duke,\TUNL]{J.~Runge,}
\author[\IU]{D.J.~Salvat,}
\author[\CMU]{A.M.~Salyapongse,}
\author[\Duke]{K.~Scholberg,}
\author[\Mephi]{A.~Shakirov,}
\author[\Mephi,\ITEP]{G.~Simakov,}
\author[\Duke,2]{G.~Sinev,}\note{Now at: South Dakota School of Mines and Technology, Rapid City, SD, 57701, USA}
\author[\IU]{W.M.~Snow,}
\author[\Mephi]{V.~Sosnovstsev,}
\author[\IU]{B.~Suh,}
\author[\IU]{R.~Tayloe,}
\author[\VT]{K.~Tellez-Giron-Flores,}
\author[\IU,3]{I.~Tolstukhin,}\note{Now at: Argonne National Laboratory, Argonne, IL, 60439, USA}
\author[\NCCU,\TUNL]{E.~Ujah,}
\author[\IU]{J.~Vanderwerp,}
\author[\ORNL]{R.L.~Varner,}
\author[\Laurentian]{C.J.~Virtue,}
\author[\IU]{G.~Visser,}
\author[\Tufts]{T.~Wongjirad,}\Tuftsdesc
\author[\CMU]{Y.-R.~Yen,}
\author[\SNU]{J.~Yoo,}\SNUdesc
\author[\ORNL]{C.-H.~Yu,}
\author[\IU,4]{J.~Zettlemoyer}\note{Now at: Fermi National Accelerator Laboratory, Batavia, IL, 60510, USA}

\emailAdd{a\_konovalov@itep.ru}

\abstract{We present results of several measurements of CsI[Na] scintillation response to 3-60 keV energy nuclear recoils performed by the COHERENT collaboration using tagged neutron elastic scattering experiments and an endpoint technique. Earlier results, used to estimate the coherent elastic neutrino-nucleus scattering (CEvNS) event rate for the first observation of this process achieved by COHERENT at the Spallation Neutron Source (SNS), have been reassessed. We discuss corrections for the identified systematic effects and update the respective uncertainty values. The impact of updated results on future precision tests of CEvNS is estimated. We scrutinize potential systematic effects that could affect each measurement. In particular we confirm the response of the H11934-200 Hamamatsu photomultiplier tube (PMT) used for the measurements presented in this study to be linear in the relevant signal scale region.}

\keywords{CsI[Na], scintillation, nuclear recoils, quenching factor, coherent elastic neutrino-nucleus scattering}

\arxivnumber{2111.02477} 

\collaboration[c]{(The COHERENT Collaboration)}

\begin{document}
\maketitle
\flushbottom


\section{Introduction} \label{sec:intro}

Characterization of various sensitive media response to ionizing radiation is a big part of the global particle detector instrumentation effort. Both new and well-known materials are being studied to expand the area and to increase the precision of the feasible experimental research. Close attention is still given to inorganic crystal scintillators as a detector type providing a good balance of signal yield, density, mechanical properties and cost. It is known that only a fraction of energy deposited in a scintillator is converted into scintillation light while the remaining part is lost to ionization and heat. Further, the light response depends on the type of incident particle through energy dissipation mechanisms within the medium. Electron recoils (ER) from gamma or beta depositions produce larger signals than nuclear recoils (NR) of the same energy induced by neutrons, by coherent elastic neutrino-nucleus scattering (CEvNS) \cite{Freedman_1974, Frankfurt_1974} or by hypothetical weakly interacting massive particles. This disparity is described by the quenching factor (QF) value --- the ratio of light yield for NR events to that of ER events of the same energy deposition. The main source of information about the QF of inorganic scintillators and its dependence on the NR energy is by measurement. There is no reliable calculation of the QF from first principles. An attractive semi-empirical model \cite{Tretyak_2010} based on Birks' law \cite{Birks_1951} and simulations of the stopping power \cite{ESTAR,SRIM} was a good description of QF data available at the time of its publication, but this model is in contradiction with low NR energy behavior of QF observed in recent measurements (e.g. ref. \cite{Joo_2019,Collar_2021}). Both modifications of the model \cite{Collar_2019} and refinements of energy-transfer mechanism understanding \cite{Bringa_2002} are discussed in the literature.

The NR QF measurements are performed via dedicated neutron calibrations of the material under study (scatterer). In a tagged elastic neutron scattering approach, the energy of a nuclear recoil $\Delta E$ is identified by the incident neutron beam energy $E_n$ and the scattering angle of the neutron $\theta$ according to Equation \ref{eq:es1} \cite{Arneodo_2000},

\begin{equation}
  \label{eq:es1}
  \Delta E=\frac{2E_n M^2_n}{(M_n+M_T)^2}\left[\frac{M_T}{M_n}+\sin{^2\theta} - \cos{\theta} \sqrt{ \left( \frac{M_T}{M_n} \right)^2- \sin{^2\theta} }  \right],
\end{equation}
where $M_n$ is the mass of the incident neutron and $M_T$ is the mass of the target nucleus. In the limit of $M_T \gg M_n$ one can simplify this expression to

\begin{equation}
  \label{eq:es2}
  \Delta E\approx\frac{2E_n M_n M_T}{(M_n+M_T)^2} (1-\cos{\theta}).
\end{equation}
The scattering angle $\theta$ can be determined by the position of a backing detector (BD) capable of the gamma/neutron pulse shape discrimination (PSD). The signals from the scatterer acquired in coincidence with neutron-like signals of the BD represent the response of the studied material to nuclear recoils of certain energy. In a QF measurement performed using the endpoint approach \cite{Joshi_2014}, the continuous spectrum of NR energy depositions from all neutron scattering angles is observed. The maximal (``endpoint'') NR energy deposition corresponds to the case of $\theta=180^{\circ}$ from Eq.~\ref{eq:es2}. The tagged neutron and the endpoint approaches are subject to different systematic effects. The first method selects a narrow region of NR energies, but assumes that such a selection can cleanly and reliably be made. The second method allows a measurement of the maximal NR energy deposition, but the data require careful treatment of the spectral shape including contributions from lower-energy NR, multiple scattering of neutrons and inelastic neutron scattering with a gamma escape.

A QF measurement also requires evaluation of the ER signal scale in the studied material. It may be a challenging task to measure the ER response at the exact same energies as those of NRs from the neutron calibrations. Rather, often the light yield of conventional gamma calibration lines such as 59.5~keV, 122~keV or 662~keV of $^{241}$Am, $^{57}$Co and $^{137}$Cs respectively is used to establish the ER scale and to define the QF value. In this case it is important to keep in mind that a scintillator's response to the ERs can depend on the ER energy \cite{Aitken_1967, Mengesha_1998, Beck_2015, Salakhutdinov_2015} so that one must scale QF values established for the different gamma lines appropriately for comparison.

In this study we focus on the response of CsI[Na] to low energy nuclear recoils. This inorganic scintillator was used by the COHERENT collaboration for the first observation of CEvNS at the SNS at Oak Ridge National Laboratory \cite{Akimov_2017} (further referred to as the COH-CsI experiment). The tension between two dedicated COHERENT QF measurements \cite{Scholz_2017, Grayson_2017} resulted in a dominant 25\% uncertainty on the expected CEvNS event rate in COH-CsI. In this paper we scrutinize the CsI[Na] QF measurements of 2016 and present two new, subsequent QF measurements performed by COHERENT. Section \ref{sec:beam_en} is dedicated to the general description of the acquired datasets with a focus on characteristics of the neutron beams utilized. Section \ref{sec:csi_cal} presents the results of the ER energy scale calibration for each of the four datasets. In sections \ref{sec:coh1} and \ref{sec:coh2} we discuss the data analysis and the updated results of the early COHERENT-1 and COHERENT-2 QF measurements, paying special attention to the correction of the identified systematic effects. We present the data analysis and results of new COHERENT-3 and COHERENT-4 QF measurements in sections \ref{sec:coh3} and \ref{sec:coh4}. A joint fit of CsI[Na] QF data, appropriate for COHERENT CEvNS measurements by this crystal, is presented in section \ref{sec:global_fit}. Finally, section \ref{sec:share} discusses the dissemination of our data to facilitate a robust and reproducible assessment of the CsI[Na] quenching factor. We use appendices to discuss technical questions relevant to understanding and interpretation of our results. Appendix~\ref{sec:meth} describes in detail several crucial aspects of our data analysis methods. Appendix~\ref{sec:pmt} is entirely dedicated to characterization of the response linearity of H11934-200 Hamamatsu ultra bialkali photomultipler tube (UBA PMT) used in our QF studies.

\section{Experimental setup and acquired datasets} \label{sec:beam_en}

The NR QF measurements described here were performed at the Triangle Universities Nuclear Laboratory's (TUNL) Tandem Van de Graaff Generator in Durham, North Carolina. The tandem is capable of accelerating light ions to generate quasi-monochromatic neutron beams via various channels including $^7$Li($p$,$n$), D(D,$n$)$^3$He, $^3$H($p$,$n$)$^3$He and $^3$H(D,$n$)$^4$He. The beam operates in either direct current (DC) or pulsed mode with a pulse period of 400$\times$ 2$^k$ ns with integer $k\geq0$ and a pulse width of about 10 ns. The variety of neutron production channels can provide beam energies in the 50~keV--30~MeV range. This facility was previously used to evaluate the NR QF for organic scintillators such as EJ-301 \cite{Awe_2018}, EJ-228 \cite{Weldon_2020_EJ}, EJ-260 \cite{Awe_2020}, stilbene \cite{Weldon_2019, Weldon_2020_SB} and inorganic sensitive media such as two-phase xenon \cite{Lenardo_2019} and neon gas \cite{Balogh_2021}.

All QF measurements presented in this work were performed using a 14.5 cm$^3$ cylindrical CsI[Na] crystal with a length of 51 mm and a diameter of 19 mm acquired from Proteus Inc. \cite{Proteus}. This crystal was grown with the same method and dopant concentration, 0.114 mol\%, as the large 14.6 kg CsI[Na] detector used for the first CEvNS measurement at the SNS \cite{Akimov_2017}. The same small crystal was used for the QF measurements presented in ref.~\cite{Collar_2019}. All COHERENT measurements at TUNL utilised the same Hamamatsu H11934-200 PMT unit with ultra-bialkali (UBA) photocathode \cite{H11934} to read out this small CsI[Na] detector. In this work we discuss four CsI[Na] QF measurements summarized in Table \ref{Table_Data}. The COHERENT-1 and COHERENT-2 measurements were conducted in 2016. The QF value estimate of 8.78$\pm$1.66\% for 5-30 keV nuclear recoil energy was evaluated based on results of these and used to predict the CEvNS rate for ref. \cite{Akimov_2017}. The large 19\% uncertainty of the QF value and respective 25\% uncertainty on the CEvNS rate was driven by the discrepancy between COHERENT-1/2 results and affected the precision of the first CEvNS measurement. The COHERENT-3 measurement was performed as a check of COHERENT-1/2 results near the largest nuclear recoil energy of $\sim$17 keV$_{nr}$ probed by both measurements. COHERENT-1/2/3 were performed using a tagged elastic neutron scattering approach. The COHERENT-4 measurement used an endpoint approach. The COHERENT-4 measurement did not rely on coincidences with BD signals, and thus checks for potential bias related to the selection applied to the BD signals.

\begin{table}
\begin{center}
\parbox{0.9\linewidth}{\caption{\label{Table_Data} A summary of COHERENT CsI[Na] QF measurements and naming conventions, compared to historical naming used in previous publications.}}
\begin{tabular}{llllll}
\hline
Dataset    & Name in \cite{Akimov_2017} & Name in \cite{Collar_2019} & Acquired & Method & Reference \\ \hline
COHERENT-1 &  COHERENT (Duke)    & Duke      & Jan.-Feb. 2016 & Tagged $n$ &  \cite{Akimov_2017,Grayson_2017} \\
COHERENT-2 &  COHERENT (Chicago) & Chicago-2 & Jan. 2016 & Tagged $n$  &  \cite{Akimov_2017,Scholz_2017} \\
COHERENT-3 &  n/a               & n/a       & Apr. 2018 & Tagged $n$  &  this work \\
COHERENT-4 &  n/a               & n/a       & Dec. 2017 & Endpoint  &  this work
\end{tabular}
\end{center}
\end{table}

\begin{figure}[ht]
\centering
\includegraphics[width=1.0\textwidth]{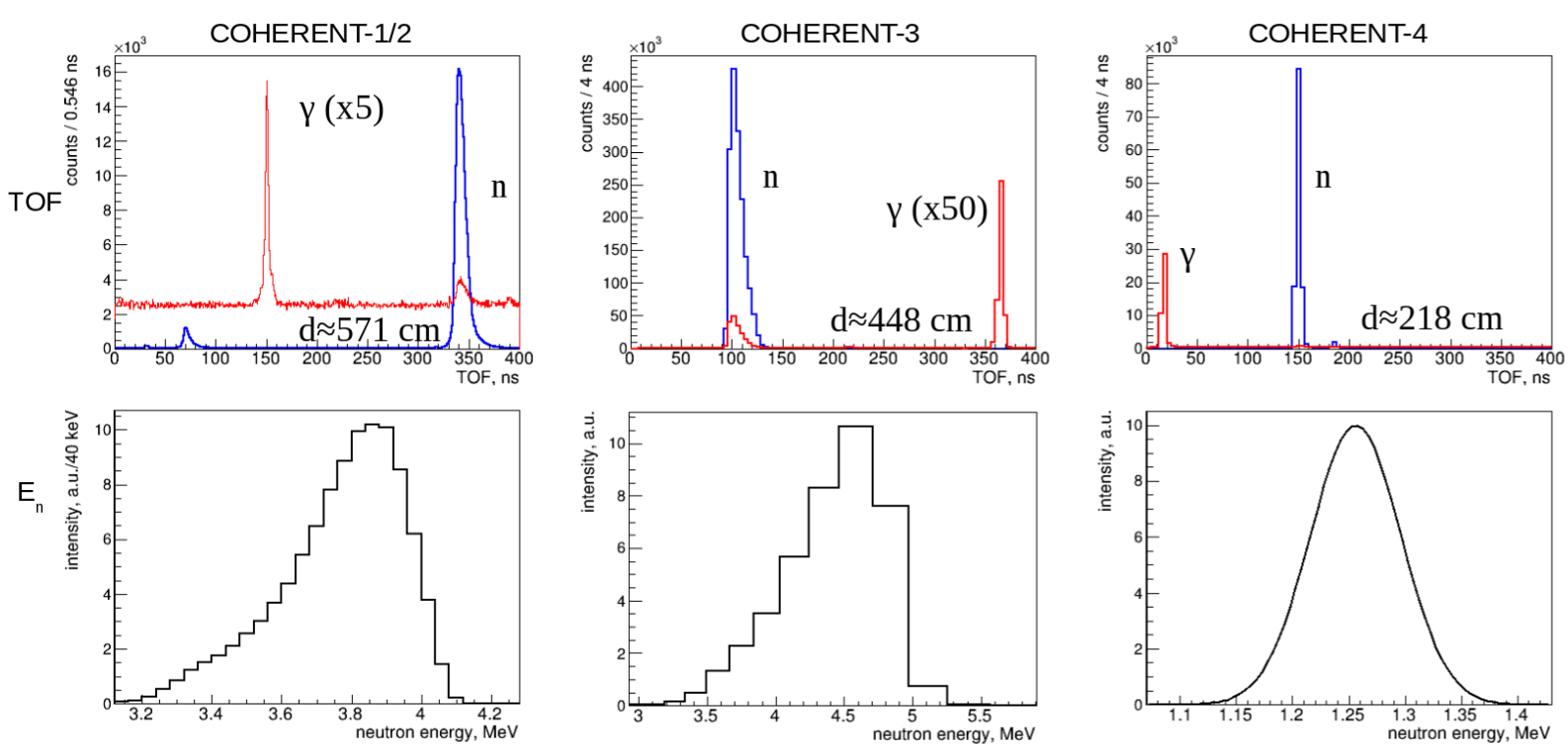}
\caption{\label{fig:tof} The TOF and neutron energy distributions for COHERENT CsI[Na] QF measurements. Top row -- TOF defined relative to the periodic BPM signal with gamma-induced BD signals marked in red and neutron-induced in blue, "d" is for the EJ-309 cell stand-off distance. Bottom row -- the neutron energy distributions for COHERENT-1/2 (left), COHERENT-3 (middle) and COHERENT-4 (right). The neutron energy distributions for COHERENT-1/2/3 were recalculated from the TOF, while COHERENT-4 is represented by a Gaussian model --- its FWHM is of the order of TOF resolution.}
\end{figure}

COHERENT-1/2/3 used the beamline associated with the ``Shielded Source Area" and a D(D,$n$)$^3$He reaction (Q-value of +3269 keV \cite{Qcalc,AME_2016}). The neutron beam energy distribution was evaluated based on the neutron time-of-flight (TOF) measurements taken before the QF experiment. For this purpose, an EJ-309 liquid scintillator detector collected data in the neutron beamline at three different stand-off distances from the neutron production target, together with beam timing information provided by a periodic beam-pickoff monitor (BPM) signal. To evaluate the neutron beam energy, we compared TOF of neutron populations relative to that of gamma events observed for different stand-off distances. The COHERENT-4 endpoint measurement used the $^7$Li($p$,$n$)$^7$Be reaction (Q-value of -1644 keV) as a neutron source. Apart from the main neutron production channel, the observed TOF distribution suggests an additional contribution from the $^7$Be$^*$ excited state providing 5-10\% of total neutron flux and with energy smaller by $\sim$ 500 keV relative to the ground state mechanism \cite{Meadows_1972,Liskien_1975}. The accuracy of our neutron beam energy estimates is limited by the TOF data sampling as well as uncertainties in the neutron production site within the source and uncertainties in the neutron interaction site within the EJ-309 detector. Examples of TOF and beam energy distributions are shown in Figure~\ref{fig:tof} and a neutron beam energy summary is presented in Table~\ref{Table_Beam}.

\begin{table}
\begin{center}
\parbox{0.85\linewidth}{\caption{\label{Table_Beam} Mean neutron beam energy in COHERENT CsI[Na] QF measurements, along with the width and error on the mean of the neutron energy distribution.}}
\begin{tabular}{lllll}
\hline
Dataset    & n source & n energy mean, MeV & FWHM, MeV & Uncertainty \\ \hline
COHERENT-1 &  D(D,$n$)$^3$He & 3.8         & 0.4  &  2\% \\
COHERENT-2 &  D(D,$n$)$^3$He & 3.8         & 0.4  &  2\% \\
COHERENT-3 &  D(D,$n$)$^3$He & 4.4         & 0.8  &  4\% \\
COHERENT-4 &   $^7$Li($p$,$n$)$^7$Be & 0.94 / 1.26 & 0.1 / 0.1  &  5\% / 4\%
\end{tabular}
\end{center}
\end{table}

\section{Calibration of CsI[Na] energy scale} \label{sec:csi_cal}

\begin{figure}[ht]
\centering
\includegraphics[width=1.0\textwidth]{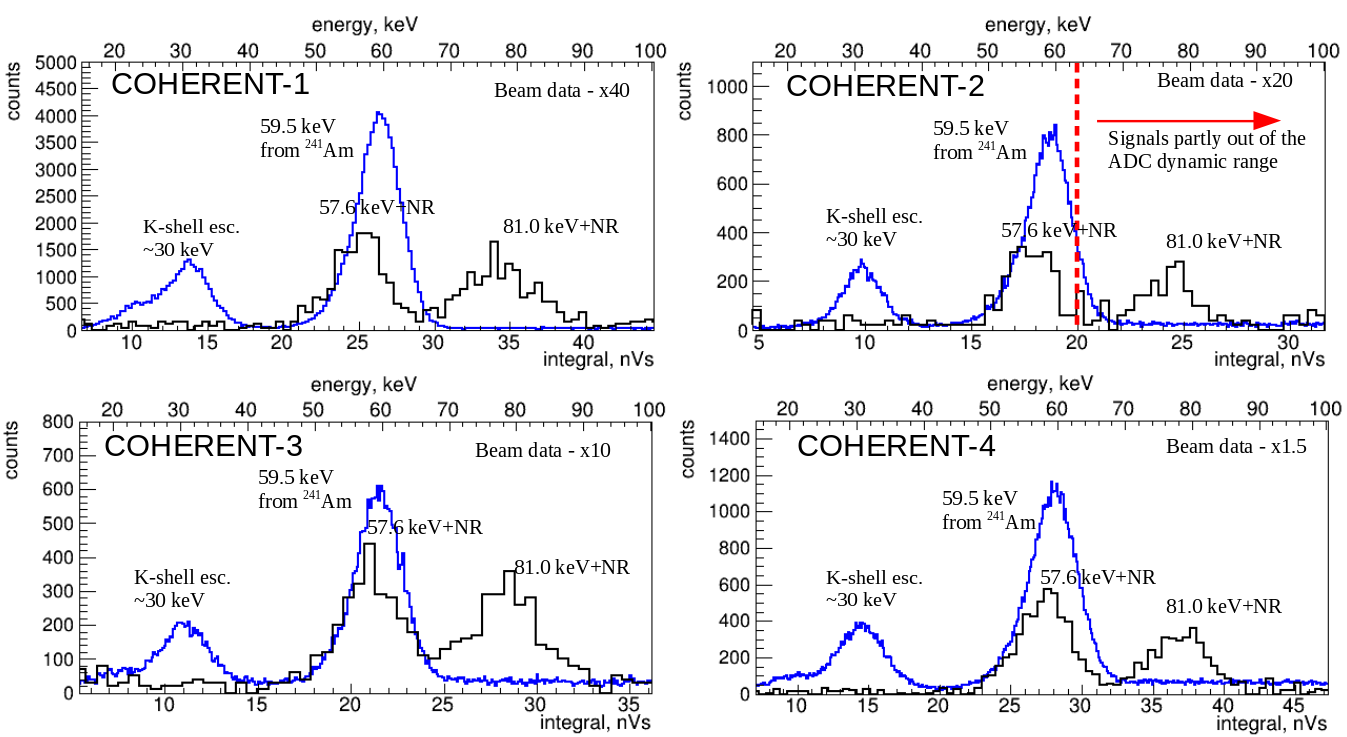}
\caption{\label{fig:csi_cal} Calibration of CsI[Na] ER energy scale in $^{241}$Am runs (blue) and $(n,n'\gamma)$ neutron beam data (black) in COHERENT QF measurements. The $(\gamma,\gamma')$ spectra are omitted for clarity. The top energy scale is calibrated by 59.5~keV and does not take into account the non-linearity of CsI[Na] crystal light response. The COHERENT-1 signal energy scale in nVs was scaled down 10 times to account for the amplification of a PS-771 used in the measurement.}
\end{figure}

In all COHERENT measurements, we use the 59.5~keV line of $^{241}$Am as a reference point to calibrate the ER light yield. The calibration runs were acquired without shaping electronics using an amplitude-triggered analog-to-digital converter (ADC) to digitize and record CsI[Na] signal waveforms. The analysis software inspected each waveform to find the start of the largest PMT pulse in the trigger region, which was utilised as a signal onset. The signal was then integrated for 3~$\mu$s from the onset. This onset finding approach provides a robust onset estimate above $\sim$30~keV which was confirmed by simulations and inspection of multiple waveforms. The NR signals in the neutron beam data were integrated for the same 3~$\mu$s starting from the first photoelectron (PE) pulse found in the region of interest. Integration and onset identification approaches were identical to the strategy COHERENT used for the first CEvNS detection using the CsI[Na] detector at the SNS~\cite{Akimov_2017}. We verify the ER calibration in the neutron beam data by looking at the gamma rays of 57.6 keV from $^{127}$I($n$,$n'\gamma$), $^{127}$I($\gamma$,$\gamma'$) \cite{Hashizume_2011} and 81.0 keV from $^{133}$Cs($n$,$n'\gamma$), $^{133}$Cs($\gamma$,$\gamma'$)\cite{Khazov_2011} reactions in coincidence with the beam spill. The ($n$,$n'\gamma$) energy depositions contain a nuclear recoil along with the gamma ray of the nominal energy. The COHERENT-1/2 ER energy verification utilized triggers in coincidence with BDs at angles restricting the NR contribution to be lower than 12~keV$_{nr}$. Such a selection allows suppression of uncertainty of the NR contribution to the total energy release. The energy deposition spectra from the calibration runs and in-situ beam data analysis are shown in Figure \ref{fig:csi_cal}. Note that "nVs" scale may slightly differ between measurements due to the different cable lengths between the PMT and ADCs.

The 59.5~keV $^{241}$Am peak is asymmetric with a more pronounced tail at lower integral. This shape is observed in all of COHERENT CsI[Na] data including the 14.6~kg CsI[Na] crystal deployed at SNS and in results from different authors. Originally \cite{Akimov_2017,Scholz_2017,Grayson_2017}, it was assumed that this asymmetry arises from the contribution of L-shell escape with energy deposition of $\sim$~55~keV.  But, this would require a similar number of K-shell and L-shell escape events which is not viable given the difference in escape gamma energies (29~keV vs. 5~keV respectively) and probability to escape. Further, MCNP6-based and Geant4-based simulations do not suggest a significant L-shell contribution, but rather partial energy loss of the 59.5 keV in inactive materials surrounding the crystal. These two different interpretations of the spectral shape only affect the crystal response by 1\% at the 59.5~keV line, but we mention it for completeness. We also mention that the DAQ in the COHERENT-2 measurement was designed so that signals of $\lesssim$~60~keV fit entirely in the digitization range of the ADC, but higher energy depositions are biased. The results of the ER energy scale calibration and verification are summarized in Table \ref{Table_LY}. The relative light yield estimates presented in the table incorporate the NR contribution for ($n$,$n'\gamma$) reactions. The uncertainties of the relative light yield take into account the accuracy of the PMT gain matching between the calibration and beam data as well as uncertainties of the total absorption peak positions within the spectra. While the relative light yield value expected for the 57.6 keV is 1.0, the 81.0 keV is affected by the CsI[Na] light response non-linearity. The ratio of light yields at 81.0 keV and 59.5 keV~is~expected~to~be~$0.98\pm0.02$~\cite{Mengesha_1998, Salakhutdinov_2015, Beck_2015}~(see~also~Figure~\ref{fig:rel_ly}).

The ER energy scale evaluated with our $^{241}$Am calibration is consistent with in-situ inelastic lines within 3\% --- close to the accuracy of the measurements. We also note that results from the analysis of $(n,n'\gamma)$ and $(\gamma,\gamma')$ subselections differ by about 3\%. Such a discrepancy may be caused by the details of selections applied to the data or presence of unresolved low intensity lines in the spectral substrate. The estimates of an absolute light yield vary from 16.5~PE/keV for \mbox{COHERENT-2} to 14.0~PE/keV for COHERENT-1 and $\sim11.5$~PE/keV for COHERENT-3/4. This discrepancy motivated the H11934-200 PMT non-linearity claim described in ref. \cite{Collar_2019}. We scrutinize this claim in Appendix~\ref{sec:pmt}. The difference between light yield estimates in \mbox{COHERENT-1/2} performed weeks apart from each other may be related to modified optical coupling (optical grease) between the crystal and the PMT. Lower light yield values in data taken in 2017/2018 compared to 2016 data suggests aging of the photocathode. It is also shown in Appendix~\ref{sec:pmt} that use of a Gaussian single photoelectron (SPE) model can bias the H11934-200 mean SPE charge estimate and affect the corresponding absolute light yield estimate by about 10\%. We emphasize that the absolute light yield value uncertainty related to SPE charge spectrum fit model does not affect evaluated QF values to leading order, as corresponding SPE mean charge values are cancelled out in the QF definition.

\begin{table}
\begin{center}
\parbox{0.85\linewidth}{\caption{\label{Table_LY} Absolute light yield (LY) estimates at 59.5 keV based on a Gaussian single photoelectron (SPE) model; light yield estimates at 57.6 keV and 81.0 keV relative to 59.5~keV (Rel. LY).}}
\begin{tabular}{ccccccc}
\hline
Dataset  & BV, V\footnotemark & LY, PE/keV & \multicolumn{2}{c}{Rel. LY (57.6 keV)} & \multicolumn{2}{c}{Rel. LY (81.0 keV)}\\
& & & $(n,n'\gamma)$ & $(\gamma,\gamma')$ & $(n,n'\gamma)$ & $(\gamma,\gamma')$ \\
\hline
COHERENT-1 & -950 & $14.0\pm0.2$ & $0.97\pm0.02$ & $1.01\pm0.02$ & $0.95\pm0.02$ & $0.97\pm0.02$\\
COHERENT-2 & -935 & $16.5\pm0.3$ & $0.97\pm0.02$ & $1.00\pm0.02$ & $0.95\pm0.02$ & $0.97\pm0.02$\\
COHERENT-3 & -[950,990] & $11.5\pm0.3$ & $0.98\pm0.03$ & $1.03\pm0.03$ & $0.95\pm0.03$ & $0.98\pm0.03$\\
COHERENT-4 & -[980,990] & $11.2\pm0.3$ & $0.99\pm0.03$ & $1.00\pm0.03$ & $0.96\pm0.03$ & n/a\footnotemark
\end{tabular}
\end{center}
\end{table}

\footnotetext{Exact PMT bias voltage (BV) was not logged for COHERENT-3/4 measurements. For each measurement the PMT was biased to a single value of voltage from the intervals in the table; the value was not changed during a measurement.}

\footnotetext{The 81.0 keV line is unresolved in the subselection of COHERENT-4 neutron beam data coincident with the time of flight characteristic for the gamma ray population.}

\section{COHERENT-1 measurement} \label{sec:coh1}

The COHERENT-1 measurement was performed in 2016 using a 3.8-MeV D(D,$n$)$^3$He neutron beam. Twelve EJ-309 backing detectors were used to tag neutrons scattering in the CsI[Na] crystal. The crystal was read out by a H11934-200 PMT held at a -950~V bias voltage.  Its signal was amplified 10 times by a PS771 amplifier with output sent to a 14-bit, 14-channel, 500 MS/s CAEN V1730 ADC. The timing of events relative to the beam pulse was determined by digitizing the signal from the BPM system. All signals from each BD were split. One copy was used to trigger the ADC by sending the signal to a channel of a MPD-4 pulse shape discriminator module. This module generated a logical signal indicating the presence of a neutron-like event in a BD. The second copy of the BD signals went directly into the ADC with a dedicated channel for each BD. Each recorded waveform was 30 $\mu$s long with a trigger occurring at $\sim$~5.5~$\mu$s. A more detailed description of the experimental setup, electronics and data acquisition can be found in ref. \cite{Grayson_2017}.

\begin{figure}[ht]
\centering
\includegraphics[width=0.8\textwidth]{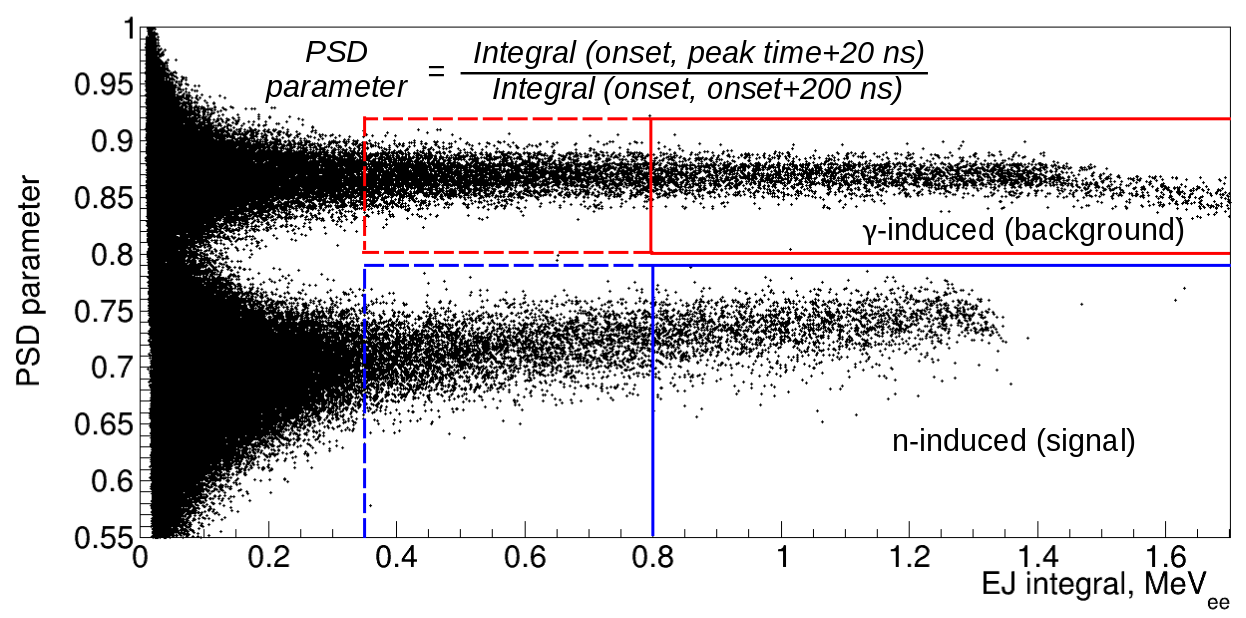}
\caption{\label{fig:coh1_psd} PSD parameter vs. integral of signals in the BD located at the 55.8$^{\circ}$ neutron scattering angle. Blue and red lines show subselections of signal and background events respectively. Dashed lines show regions with softer BD integral threshold close to the one used in the original analysis \cite{Grayson_2017}. The BD thresholds were increased in the consequent analysis cross-check to suppress contribution of inelastic events (see text).}.
\end{figure}

\begin{figure}[ht]
\centering
\includegraphics[width=1.0\textwidth]{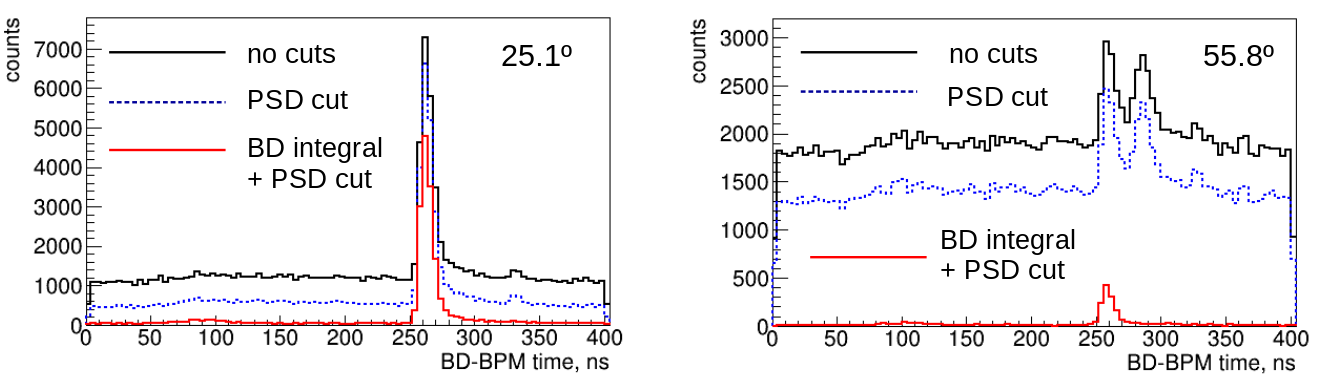}
\caption{\label{fig:coh1_tof} TOF characteristic parameter calculated based on the BD-BPM time separation for the BD detectors at  25.1$^{\circ}$~(left) and 55.8$^{\circ}$~(right) neutron scattering angles. The effects of consequent application of PSD parameter<0.78 and the BD integral cut (specified in the Table \ref{Table_COH1}) are shown. The second peak in the plot for the 55.8$^{\circ}$ angle consists of events with the neutron-like PSD parameter, but with a soft BD integral spectrum corresponding to neutrons which either scatter inelastically in the CsI[Na] or lose energy in the surrounding materials prior to energy release in the BD.}.
\end{figure}

\begin{figure}[ht]
\centering
\includegraphics[width=1.05\textwidth]{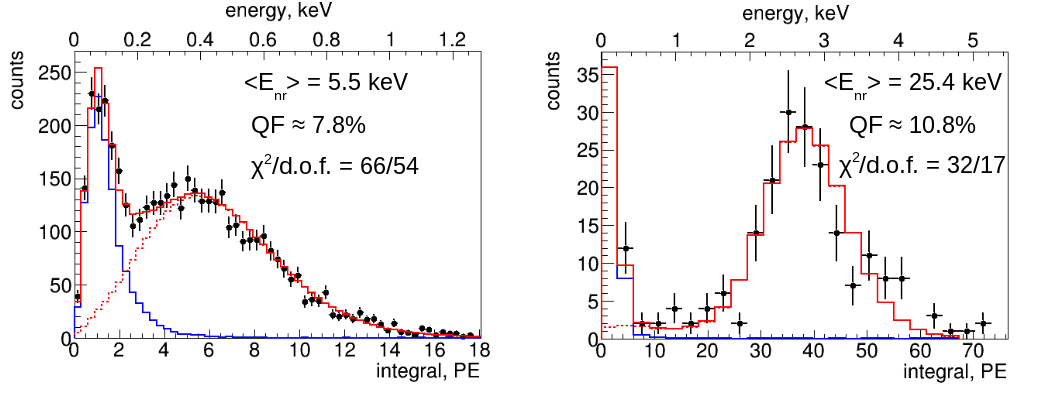}
\caption{\label{fig:coh1_nr}  An illustration of the NR spectra from the CsI[Na] detector, best fit overlaid (solid red line). NR and background contributions are shown as dashed red and solid blue lines respectively. Left: 25.1 degrees scattering angle. Right: 55.8 degrees scattering angle.}.
\end{figure}

We select NR signals associated with the beam by cutting on BD signal PSD parameter and integral, as well as restricting a TOF characteristic calculated by the time delay between BD and BPM signals. The illustration of the analysis parameter space which can be used for such a selection can be found in Figures~\ref{fig:coh1_psd} and \ref{fig:coh1_tof}. The spectra of NR signal integrals were fit to the prediction based on a MCNPX-Polimi simulation varying signal and background amplitudes as well as the quenching factor value. The simulation utilized the results from ref.~\cite{Pino_2014} to convert the energy of EJ-309 recoil protons to ER energy equivalent. Carbon recoil energy was converted using the same parametrization, but with an additional quenching of 20\%. A representative background PDF was obtained from the subselection of gamma-like BD signals with a TOF characteristic of the off-beam region populated by environmental gamma interactions. The results of the fits are summarized in Table~\ref{Table_COH1} and illustrated in Figure~\ref{fig:coh1_nr}.

\begin{table}
\begin{center}
\parbox{0.85\linewidth}{\caption{\label{Table_COH1} Results of COHERENT-1 CsI[Na] QF measurement}}
\begin{tabular}{p{1.0cm}p{1.6cm}p{1.6cm}p{1.6cm}ccc}
\hline
\centering Angle, deg. & E$_{nr}$ mean, \centering keV & E$_{nr}$ RMS, \centering keV & \centering BD cut, keV$_{ee}$ & E$_{vis}$, keV$_{ee}$ & QF, \% (this work) & QF, \% (ref. \cite{Grayson_2017}) \\ \hline
19.9 & 3.5 $\pm$ 0.1 &\centering 0.4 &\centering 270 & 0.25 $\pm$ 0.03  & 7.2 $\pm$ 0.8 & 7.38 $_{-0.14}^{+0.10} $\\
25.1 & 5.5 $\pm$ 0.1 &\centering\centering 0.5 &\centering 310  & 0.43 $\pm$ 0.02 & 7.8 $\pm$ 0.4 & 7.97 $_{-0.12}^{+0.13}$\\
31.3 & 8.4 $\pm$ 0.2 &\centering\centering 0.7 &\centering 330 & 0.76 $\pm$ 0.03 & 9.1 $\pm$ 0.4 & 9.32 $\pm$ 0.08\\
37.4 & 11.8 $\pm$ 0.2 &\centering 1.0 &\centering 280 & 1.16 $\pm$ 0.05 & 9.8 $\pm$ 0.4 & 10.10 $_{-0.10}^{+0.09}$\\
44.0 & 16.2 $\pm$ 0.3 &\centering 1.4 &\centering 600 & 1.65 $\pm$ 0.06 & 10.2 $\pm$ 0.4 & 10.34 $\pm$ 0.13\\
55.2 & 24.8 $\pm$ 0.5 &\centering 1.8 &\centering 810 & 2.65 $\pm$ 0.10 & 10.7 $\pm$ 0.4 & 11.00 $\pm$ 0.13\\
55.8 & 25.4 $\pm$ 0.5 &\centering 1.8 &\centering 800 & 2.75 $\pm$ 0.10 & 10.8 $\pm$ 0.4 & 11.04 $\pm$ 0.18\\
68.0 & 36.3 $\pm$ 0.7 &\centering 2.5 &\centering 810 & 3.75 $\pm$ 0.15 & 10.3 $\pm$ 0.4 & 10.76 $\pm$ 0.11\\
73.8 & 41.6 $\pm$ 0.8 &\centering 2.9 &\centering 810 & 4.29 $\pm$ 0.17 & 10.3 $\pm$ 0.4 & 10.75 $\pm$ 0.12\\
85.5 & 52.7 $\pm$ 1.1 &\centering 3.7 &\centering 810 & 5.14 $\pm$ 0.21 & 9.7 $\pm$ 0.4 & 10.40 $\pm$ 0.15\\
90.0 & 55.8 $\pm$ 1.1 &\centering 4.7 &\centering 600 & 5.18 $\pm$ 0.22 & 9.3 $\pm$ 0.4 & 9.75 $\pm$ 0.20\\
97.1 & 62.6 $\pm$ 1.3 &\centering 5.2 &\centering 580 & 5.82 $\pm$ 0.25 & 9.3  $\pm$ 0.4 & 9.67 $\pm$ 0.23
\end{tabular}
\end{center}
\end{table}

A few small corrections have been applied to initial results from ref. \cite{Grayson_2017} after scrutinizing analysis methods. Note that these corrections are taken into account in Figure~\ref{fig:csi_cal} and values from Table~\ref{Table_LY} and Table~\ref{Table_COH1} (except for the last column citing the QF values from ref. \cite{Grayson_2017}). First, the CsI[Na] response to the 59.5~keV calibration line was underestimated by 3$\%$, due to an issue with the signal onset finding algorithm used for $^{241}$Am data analysis. This can be accounted for by directly scaling QF values obtained in ref.~\cite{Grayson_2017}. The second effect neglected in the original analysis is the contribution of accidental afterglow PE to the NR signal waveforms, discussed in Appendix~\ref{subsec:afterglow}. The mean afterglow contribution expected in the 3 $\mu$s integration window is about 0.2~PE for COHERENT-1. This value contributes less than 1\% to the CsI[Na] response to NR with energies above 16 keV, but may induce a bias up to 7\% for the lowest NR energy. Uncertainties on the QF have been updated to account for this potential bias. We have also updated the model of the CsI[Na] detector energy resolution.  The leading component of the energy resolution is the Poisson fluctuations in the number of detected PE ($N_{\text{PE}}$). It is proportional to $\sqrt{N_\text{PE}}$ and was taken into account in the original analysis. Another non-negligible contribution comes from the shape of the single photoelectron (SPE) charge response of the PMT. A relatively wide SPE distribution, with an RMS to mean ratio of $\sim$~0.5, additionally smears the integral spectra. This effect is proportional to $\sim0.5\sqrt{N_\text{PE}}$ and should be applied on the top of the Poissonian smearing in the detector response model.

We also improved the treatment of neutron inelastic events. These events often generate a gamma ray, with energies from tens to hundreds of keV. If such a gamma ray interacts within the CsI[Na] crystal the total energy deposition is out of the NR ROI due to the absence of quenching. When a gamma ray escapes, the NR energy deposition is biased lower than that expected for elastic neutron scattering. The fraction of selected events that are inelastic scattering events featuring gamma escape, after requiring BD integral larger than $\sim$300 keV, is expected to be between 3\% for a scattering angle of 19.9 degrees (3.5 keV) to 10\% at 55 degrees (25 keV) and up to 60\% at 97 degrees (65 keV of nominal elastic NR energy) based on the MCNPX-Polimi simulation. The influence of inelastic scattering is included in our fitting. To validate our treatment of these inelastic events, we suppress their rate by requiring a strict BD integral cut for larger scattering angles (see the ``BD cut'' column in Table \ref{Table_COH1}). The results of this cross check coincide with the original results to a few percent. The differences between this cross check and nominal QF values are included in the QF uncertainty. Note that for COHERENT-1 results shown in refs.~\cite{Akimov_2017, Collar_2019, Grayson_2017, Scholz_2017}, the NR energy width was propagated to the QF-value uncertainty. However, this width is accounted for in our simulation, and thus does not reflect an uncertainty on the NR mean energy. We do not propagate it to the QF uncertainty in this work.

\section{COHERENT-2 measurement} \label{sec:coh2}

The COHERENT-2 measurement, performed in 2016, utilized the D(D,$n$)$^3$He neutron beam with a mean neutron energy of 3.8 MeV. The EJ299-33A plastic scintillator with PSD capability was used as a BD to tag the scattering angle of incident neutrons. The light produced by EJ299-33A was collected by a 5-inch 9390B ET Enterprises PMT biased to -750V. The raw output of the CsI[Na] PMT, biased to -935V, was fed to channel 1 of the digitizer while the raw output of the EJ299-33A PMT was passed through a 6 dB attenuator and recorded as channel 2. For each trigger,  6~$\mu$s waveforms were recorded from both the CsI[Na] and EJ299-33A detectors with a trigger position set to 2~$\mu$s. An offset of 20~ns between these two channels was found using a $^{22}$Na source located in the midpoint distance of two detectors. The scattering angle of the neutrons was defined by changing the position of the EJ299-33A BD. A detailed description of the setup as well as a description of the original analysis can be found in the thesis \cite{Scholz_2017}.

This work presents an updated COHERENT-2 data analysis which was not able to reproduce the QF values quoted by the initial analysis effort \cite{Scholz_2017}. While the calibration of the BD and CsI[Na] energy scales in the ADC units coincide within few percent in both analyses, the integrals of the NR-scale signals of CsI[Na] are discrepant as well as the SPE characteristic charge. This difference in the analyses of low energy signals is likely due to differing treatment of fluctuations in the DC baseline voltage of the ADC. These fluctuations can significantly distort few to few tens of PE signals of interest. The final version of the analysis code used in ref.~\cite{Scholz_2017} was not available to authors of this work, while the backup version used a baseline correction approach which systematically  under-integrated low energy signals. It is not clear if this approach was utilized in the final version of the initial analysis, but the code from the backup version available to us yielded QF values close to those quoted in ref. \cite{Scholz_2017}. In our re-analysis effort we reject triggers that have large baseline fluctuations which may distort signals of interest to avoid this bias. The authors of the initial analysis do not agree with our updated results, but were not available for joint analysis effort. To address possible concerns, and to ensure transparency and reproducibility of results, we will share all raw COHERENT data used for this QF result (see section \ref{sec:share}).

\begin{figure}[ht]
\centering
\includegraphics[width=1.0\textwidth]{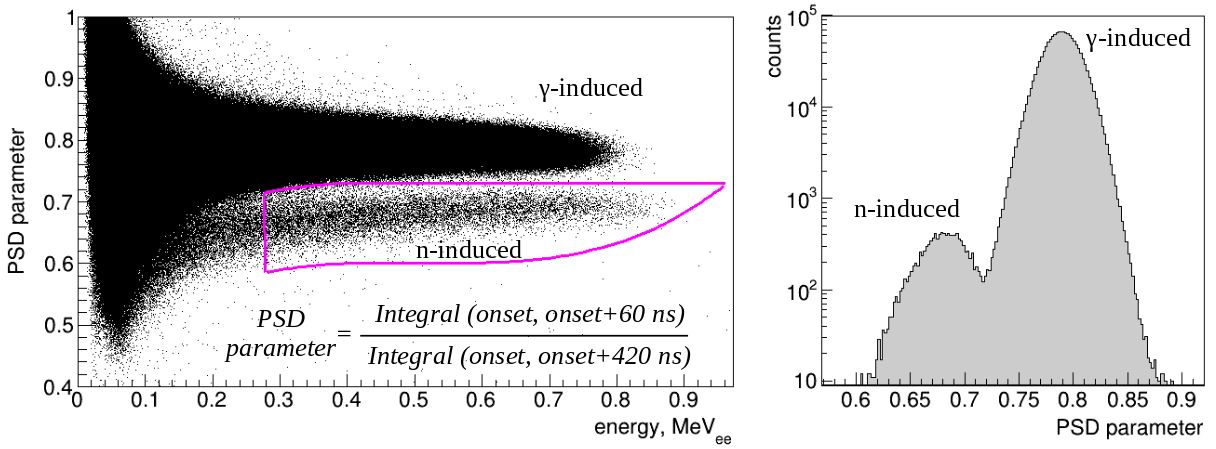}
\caption{\label{fig:coh2_psd} The PSD distribution for BD events in the neutron beam data for COHERENT-2. Left: PSD parameter vs. integral of signals in the BD.\protect\footnotemark Magenta lines show the subselection of neutron-induced interactions used to obtain NR signals for the QF analysis. Right: the PSD parameter distribution for signals with integrals larger than 280~keV$_{ee}$.}
\end{figure}

\footnotetext{Some of the n-induced BD signals exhibit the digitizer range overflow. Using the residual PSD capability for overflow events we estimate their fraction as  $\sim20\%$ above the 280 keV$_{ee}$. Following ref. \cite{Scholz_2017} we do not include overflow events in our analysis. Inclusion of these events results in the QF estimates slightly larger, but well within the respective uncertainties from Table \ref{Table_COH2}.}

\begin{figure}[ht]
\centering 
\includegraphics[width=1.0\textwidth]{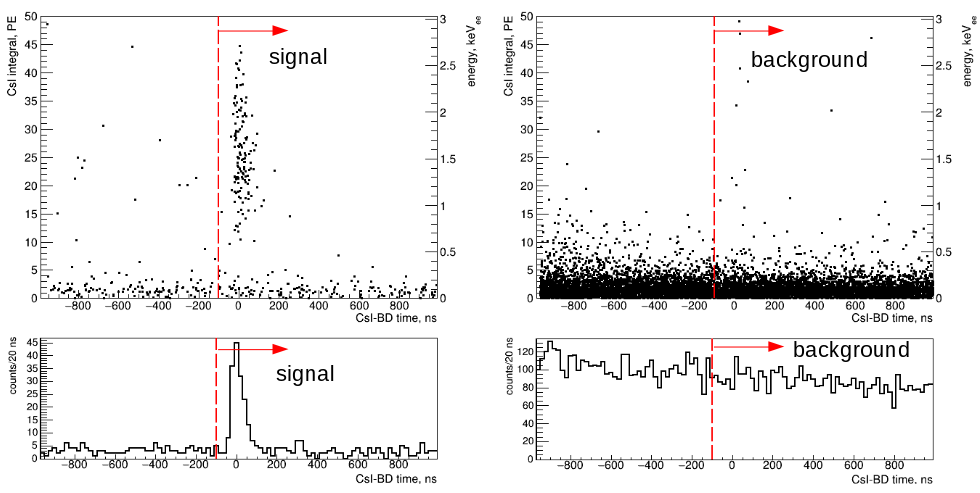}
\caption{\label{fig:coh2_tof} Time delay between CsI[Na] and BD signals in neutron-induced signal (left) and gamma-induced background (right) subselections for 45 degrees scattering angle in COHERENT-2 data. Top: CsI[Na] integral vs. time delay. Bottom: the projected time delay distributions, see discussion of the decreasing timing trend in Appendix \ref{subsec:bg_choice}. The red line marks the offset corresponding to beam-related signals.}
\end{figure}

The triggers related to neutron scattering were selected with the use of BD signals (see Figure~\ref{fig:coh2_psd}). The energy scale of the BD was evaluated based on its response to the Compton edge from 511 keV ($^{22}$Na) and 662 keV ($^{137}$Cs) gamma lines -- energy depositions of 341 and 477 keV respectively. The boundaries of a neutron-induced signal population in the PSD parameter vs. integral space were verified using neutrons from a $^{252}$Cf source. A BD integral cut of 280 keV allowed us to suppress the gamma-induced background and contributions from inelastic neutron scattering in CsI[Na]. Additional suppression of accidental backgrounds may come from the study of the relative timing between signals of CsI[Na] and the BD. A well-separated beam-related excess can be seen in Figure~\ref{fig:coh2_tof} for the triggers selected by BD PSD parameter and integral cuts from Figure~\ref{fig:coh2_psd}. The peak in the timing distribution is smeared by variation in the onset of CsI[Na] PE pulses related to scintillation decay times. The smearing becomes larger for smaller NR energies. We used only signals with a CsI[Na]--BD delay from -100 to 1000~ns to suppress accidental backgrounds. Our requirement on the end of the delay interval is loose because the first PE in CsI[Na] low energy response can arrive up to few microseconds after the interaction time. We can not extend this interval further in time since our 6~$\mu$s waveform could not accommodate a 3~$\mu$s integration window. The contribution of low PE signals arriving later than the delay allowed in our analysis was taken into account through a simulation. The fraction of events with the first PE appearing within the 1 $\mu$s after interaction in CsI[Na] is about 65\% for 1 PE, 88\% for 2 PE, 96\% for 3 PE and 99\% for 4 PE. This efficiency was considered when comparing the observed NR spectra with the prediction. We point out that the delay between an actual interaction time and appearance of an observable signal in a scintillator can be treated differently in the scope of the QF discussion. This matter is discussed in Appendix~\ref{subsec:def_lowE}.

We define a background subselection to account for contributions from random coincidence events in CsI[Na] using triggers for which the BD signal is gamma related: PSD parameter above 0.8 and BD integral larger than 280 keV$_{ee}$. These triggers are primarily caused by environmental gammas; the beam related contribution is negligible as can be seen from Figure~\ref{fig:coh2_tof}~(right). This background region contains many more triggers than the signal region so we use it as the empirically estimated background probability density function (PDF) in the fit, neglecting statistical uncertainty on the normalization. Additional information regarding choice of a background model can be found in Appendix \ref{subsec:bg_choice}.

\begin{figure}[hp]
\centering
\includegraphics[width=0.95\textwidth]{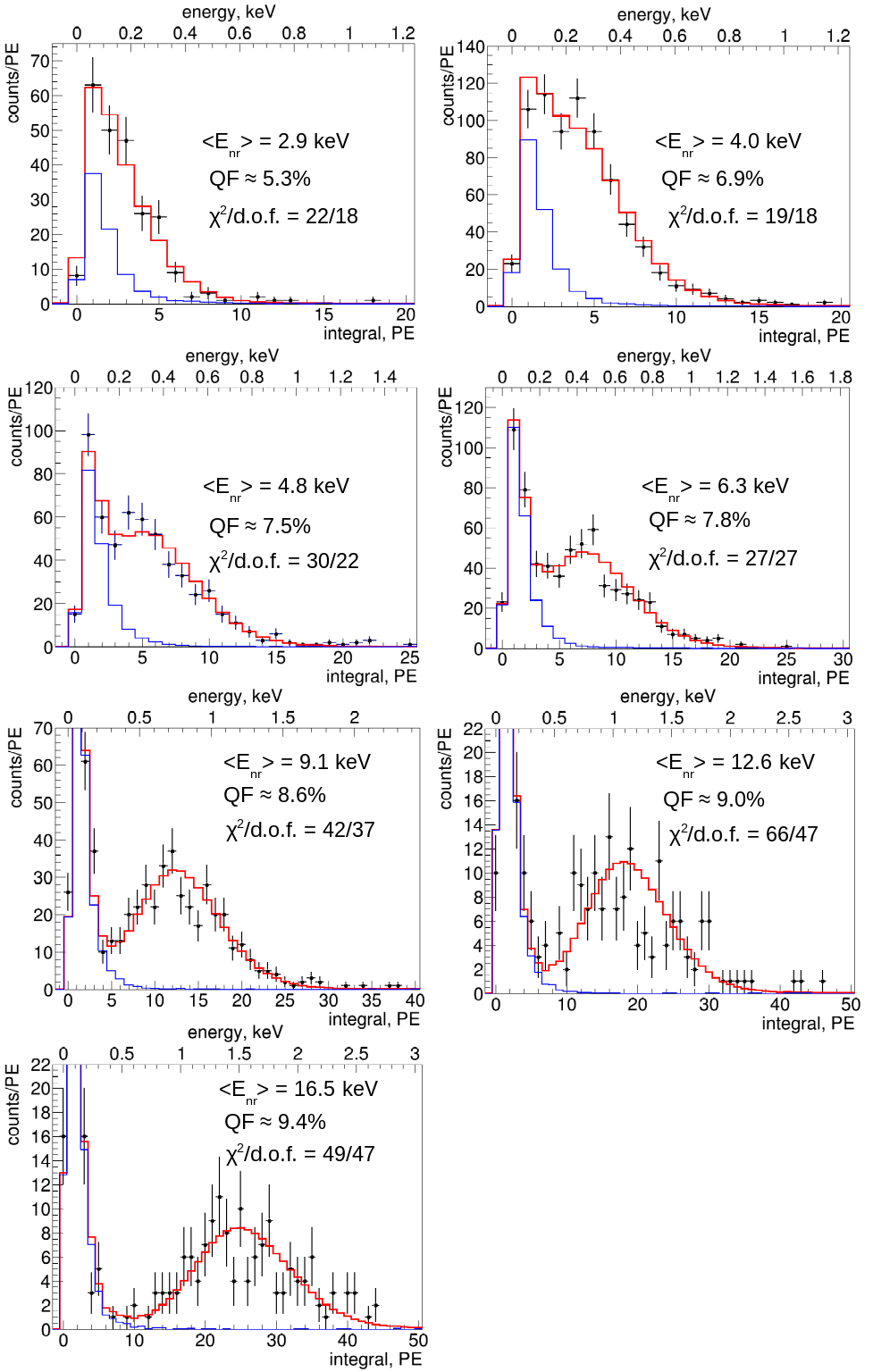}
\caption{\label{fig:coh2_nr} Experimental spectra of NR candidates in the COHERENT-2 measurement with the best-fit models overlaid: red -- total prediction, blue -- background contribution, $\chi^2/d.o.f.$ from Wilks' theorem.}
\end{figure}

We fit the CsI[Na] integral spectrum obtained with the signal selections to the sum of the signal prediction, based on the MCNPX-Polimi simulation, and the background PDF. The MCNPX-Polimi simulation of EJ299-33A utilized the modified Lindhard model parametrization \cite{Ficenec_1987,Collar_2010} of the proton recoils quenching factor and a constant quenching of 1\% for carbon recoils in accord with the original analysis \cite{Scholz_2017} of the COHERENT-2 data. Three parameters are varied in the CsI[Na] QF fit: signal amplitude, background amplitude and the QF value. For the two lowest NR energies, we fix the background component amplitudes according to the number of events in pre-beam events in the time interval [-900,-100] ns to reduce the statistical uncertainty on the measured QF. We account for the dominant sources of energy smearing: Poisson fluctuations in the number of detected photoelectrons and the width of the SPE integral distribution. The fit results are shown in Table~\ref{Table_COH2} in comparison with the initial results from ref.~\cite{Scholz_2017} and illustrated in Figure~\ref{fig:coh2_nr}. The uncertainties on E$_{vis}$ and QF include uncorrelated statistical contributions and a 4\% correlated systematic contribution from the integration approach accuracy (see Appendix~\ref{subsec:spe_int}) and ER energy scale calibration uncertainty.

\begin{table}
\begin{center}
\parbox{0.85\linewidth}{\caption{\label{Table_COH2} Results of COHERENT-2 CsI[Na] QF measurement}}
\begin{tabular}{p{1.0cm}p{2.0cm}p{1.6cm}ccc}
\hline
\centering Angle, deg. &\centering E$_{nr}$ mean, keV &\centering E$_{nr}$ RMS, keV & E$_{vis}$, keV$_{ee}$ & QF, \% (this work) & QF, \% (ref. \cite{Scholz_2017})\\
\hline
18 & 2.85 $\pm$ 0.06 &\centering 0.51 & 0.151 $\pm$ 0.023 & 5.3 $\pm$ 0.8 & 5.21  $\pm$ 1.74 \\
21 & 3.85 $\pm$ 0.08 &\centering 0.65 & 0.266 $\pm$ 0.017 & 6.9 $\pm$ 0.5 & 6.38 $\pm$ 0.76 \\
24 & 4.98 $\pm$ 0.10 &\centering 0.74 & 0.374 $\pm$ 0.021 & 7.5 $\pm$ 0.4 & 6.75 $\pm$ 0.75 \\
27 & 6.24 $\pm$ 0.12 &\centering 0.83 & 0.49 $\pm$ 0.03 & 7.8 $\pm$ 0.4 & 6.90 $\pm$ 0.65 \\
33 & 9.17 $\pm$ 0.18 &\centering 1.12 & 0.79 $\pm$ 0.04 & 8.6 $\pm$ 0.5 & 7.38 $\pm$ 0.51 \\
39 & 12.6 $\pm$ 0.3 &\centering 1.4 & 1.13 $\pm$ 0.06 & 9.0 $\pm$ 0.5 & 7.14 $\pm$ 0.67 \\
45 & 16.5 $\pm$ 0.3 &\centering 1.8 & 1.55 $\pm$ 0.08 & 9.4 $\pm$ 0.5 & 7.16 $\pm$ 0.63
\end{tabular}
\end{center}
\end{table}

\section{COHERENT-3 measurement} \label{sec:coh3}

The COHERENT-3 measurement was intended as an independent test of QF values suggested by COHERENT-1/2 results. It was performed in April 2018 using a 4.4~MeV D(D,$n$)$^3$He neutron beam. A PSD-capable EJ-309 BD was used to tag neutrons scattered at an angle of 42 degrees relative to the beam direction. Such an angle corresponds to about 17.1 keV energy of CsI[Na] nuclear recoils. We used a SIS3316 250MS/s 14 bit ADC to record signals from the raw output of the CsI[Na] PMT, the EJ-309 detector and the beam pickoff monitor (BPM). The EJ-309 signal was used to trigger the ADC. The digitization length per trigger was 40~$\mu$s for CsI[Na], 1.4~$\mu$s for the BD and 1~$\mu$s for the BPM. The pre-trigger window was set to 460~ns for the EJ-309 channel and 11~$\mu$s in the CsI[Na] channel. The exact bias voltage of the CsI[Na] PMT was not recorded, but is expected to be in the range from -950 to -990~V.

\begin{figure}[ht]
\centering
\includegraphics[width=1.0\textwidth]{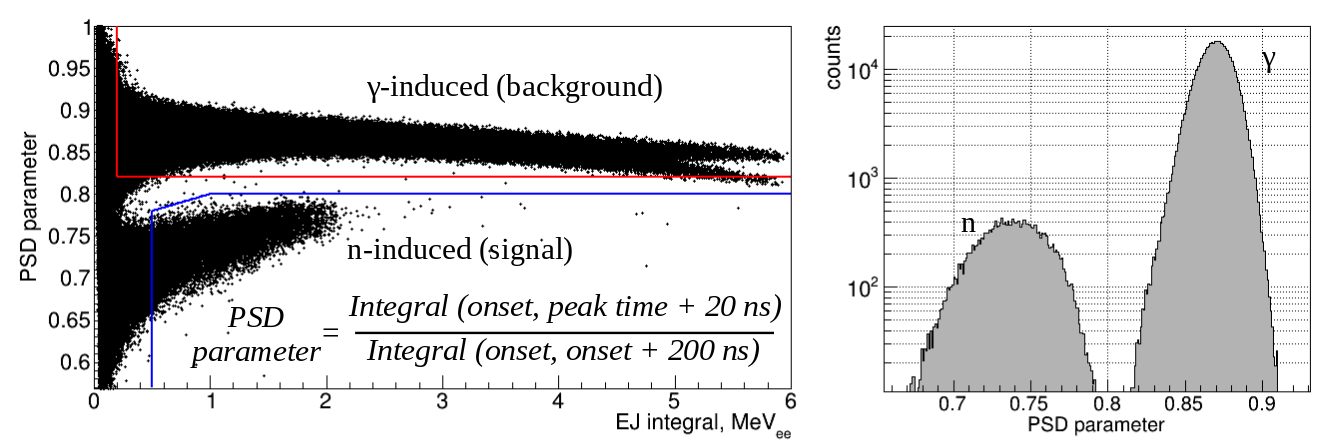}
\caption{\label{fig:coh3_psd} The PSD distribution for BD signals in neutron beam data (COHERENT-3). Left: PSD parameter vs. energy deposition for signals in the EJ-309 detector. Blue and red lines mark the selections for the neutron-induced signal events and gamma-induced background. The fork-like structure at the higher integral part is due to a combination of the ADC dynamic range edge and peculiarities of the PSD parameter definition. Right: the PSD distribution for signals with integrals larger than 0.5~MeV$_{ee}$. }
\end{figure}

\begin{figure}[ht]
\centering
\includegraphics[width=0.7\textwidth]{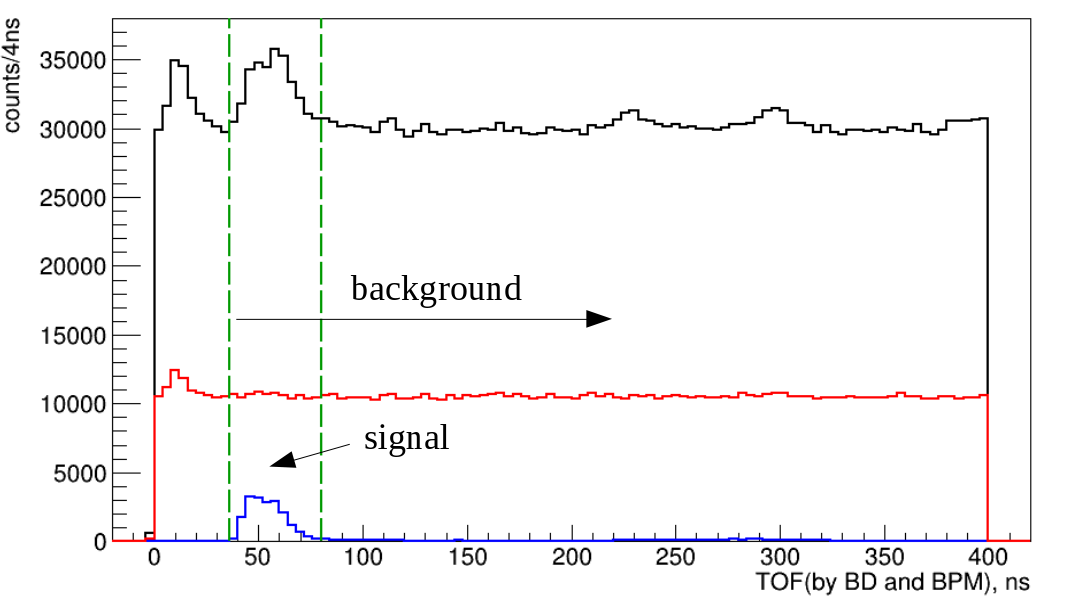}
\caption{\label{fig:coh3_tof} Time delay (TOF) between periodic BPM and BD signals. Black: before any BD cuts. Blue: events with the neutron-like PSD passing the 500 keV$_{ee}$ BD energy deposition cut. Red: events with the gamma-like PSD and the 200 keV$_{ee}$ BD energy deposition cut. The neutron-induced signal events are contained within dashed green lines. Only signals with TOF greater than 36 ns (first green line) are used to construct the background PDF to remove the beam-related excess seen at low TOF.}
\end{figure}

As in the case of COHERENT-1/2 measurements, we select neutron-induced interactions by cutting on the PSD and integral values of the BD. We also estimate the background PDF using a selection of gamma-induced triggers (see Figure \ref{fig:coh3_psd}). The BD energy scale calibration was performed using the 477 keV Compton edge of $^{137}$Cs 662 keV gammas. We use the time delay between BPM and BD signals to additionally suppress the contribution from environmental neutrons and beam neutrons that have scattered upstream in the beam-line. The distribution of this delay after imposing the PSD and BD restrictions shows a prominent beam-related excess which we select for NR candidates. We note that the background selection of gamma-like events also has a beam-related excess which we can reject to avoid bias in the background estimate (see Figure~\ref{fig:coh3_tof}). The check of CsI[Na] waveforms of the NR candidates showed that about 6\% of 10-50 PE signals have a Cherenkov-like pulse shape. We suppress them by requiring the ratio of the integral in the first 100 ns to the full 3~$\mu$s integral to be lower than 0.7 for both signal and background subselections. The efficiency of such a cut reaches $\sim100\%$ by 5 PE for regular CsI[Na] signals. We neglect inefficiency of this cut at lower integrals as this part of the NR spectrum is heavily dominated by the background. Neither COHERENT-1 nor COHERENT-2 demonstrate contributions from the Cherenkov-like signals comparable to COHERENT-3, which may be related to the specifics of the neutron irradiation setup.

\begin{figure}[hp]
\centering
\includegraphics[width=0.74\textwidth]{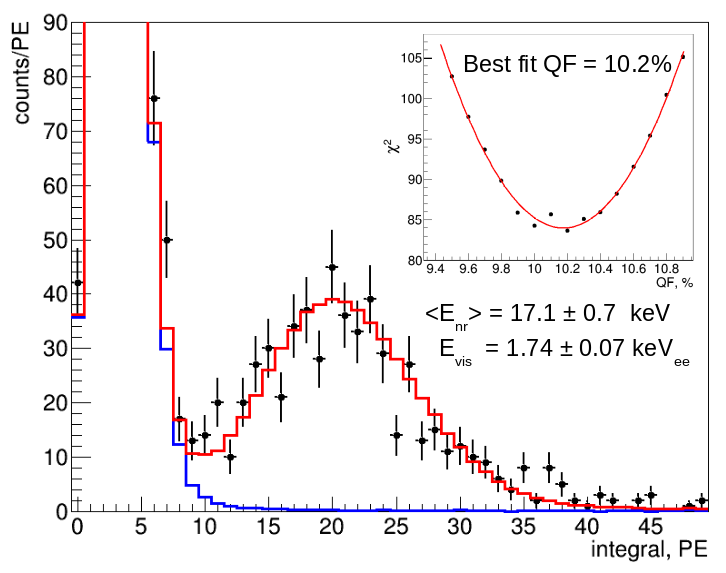}
\caption{\label{fig:coh3_nr} The NR spectrum overlaid with the best-fit prediction for the COHERENT-3 measurement. Red -- best fit corresponding to a QF of 10.2\%, blue -- best-fit background distribution. Inset: $\chi^2$ scan over fixed QF values, 46 d.o.f.}
\end{figure}

We fit the observed NR energy spectrum to a MCNPX-Polimi simulation with three parameters~--- the QF value, NR signal normalization and background normalization. The quenching of proton and carbon recoil signals in the EJ-309 simulation was treated in a fashion similar to COHERENT-1 (see Section \ref{sec:coh1}). The result of the CsI[Na] QF fit is shown in Figure \ref{fig:coh3_nr}. The best-fit QF is 10.2\% at a mean E$_{nr}$ of 17.1$\pm$0.7~keV, which corresponds to $1.74\pm0.07$ keV$_{ee}$ of visible energy (see also Table~\ref{Table_COH3}). The 4\% uncertainty on the visible energy includes statistical uncertainty as well as errors in light yield, pulse selection and integration and tests of alternative BD energy cuts. When determining the uncertainty on the QF, both the errors on the observed visible energy uncertainty and the mean E$_{nr}$ should be taken into account. The fit resulted in a modest $\chi^2/d.o.f.$ of 84/46. We performed an additional validation of the fit in which the width of the NR prediction was also allowed to float. This led to $\chi^2/d.o.f.$ closer to 1, but had no effect on the best-fit QF.

\begin{table}
\begin{center}
\parbox{0.85\linewidth}{\caption{\label{Table_COH3} Results of COHERENT-3 CsI[Na] QF measurement}}
\begin{tabular}{lllll}
\hline
Angle, deg. & E$_{nr}$ mean, keV & E$_{nr}$ RMS, keV & E$_{vis}$, keV$_{ee}$ & QF, \% \\ \hline
42 & 17.1 $\pm$ 0.7 & 1.8 & 1.74 $\pm$ 0.07 & 10.2 $\pm$ 0.6\\
\end{tabular}
\end{center}
\end{table}

\section{COHERENT-4 endpoint measurement} \label{sec:coh4}

The COHERENT-4 measurement performed in December 2017 utilized the endpoint approach to QF study. Two successive runs with 0.94 MeV and 1.26 MeV neutrons produced by a $^7$Li($p$,$n$)$^7$Be source allowed us to measure nuclear recoil spectra up to $\sim$29 keV and $\sim$39 keV endpoints respectively. We compare the observed CsI[Na] response over a range of nuclear recoil energies to several QF models given in the literature. Since there was no scattered neutron tagging, COHERENT-4 allows us to test potential systematic uncertainties related to the backing detectors used in COHERENT-1/2/3.

In the process of data acquisition the raw PMT signal was digitized with a SIS3316 250~MS/s 14 bit ADC, while another channel of the ADC recorded data from the periodic BPM (800 ns). The exact bias voltage of the H11934-200 PMT was not recorded, but it was likely in the range from -980 to -990~V. The ADC was triggered by a CsI[Na] signal with a trapezoidal filter threshold equivalent to an integral threshold of about 6~PE, reaching 100\% efficiency at about 10~PE (see pages 16-18 of ref. \cite{Struck_2014} for the trapezoidal filter description). A 40~$\mu$s waveform was recorded for the CsI[Na] channel with a 10~$\mu$s pre-trigger region, while a 1~$\mu$s waveform was recorded for the BPM channel.

\begin{figure}[ht]
\centering
\includegraphics[width=1.0\textwidth]{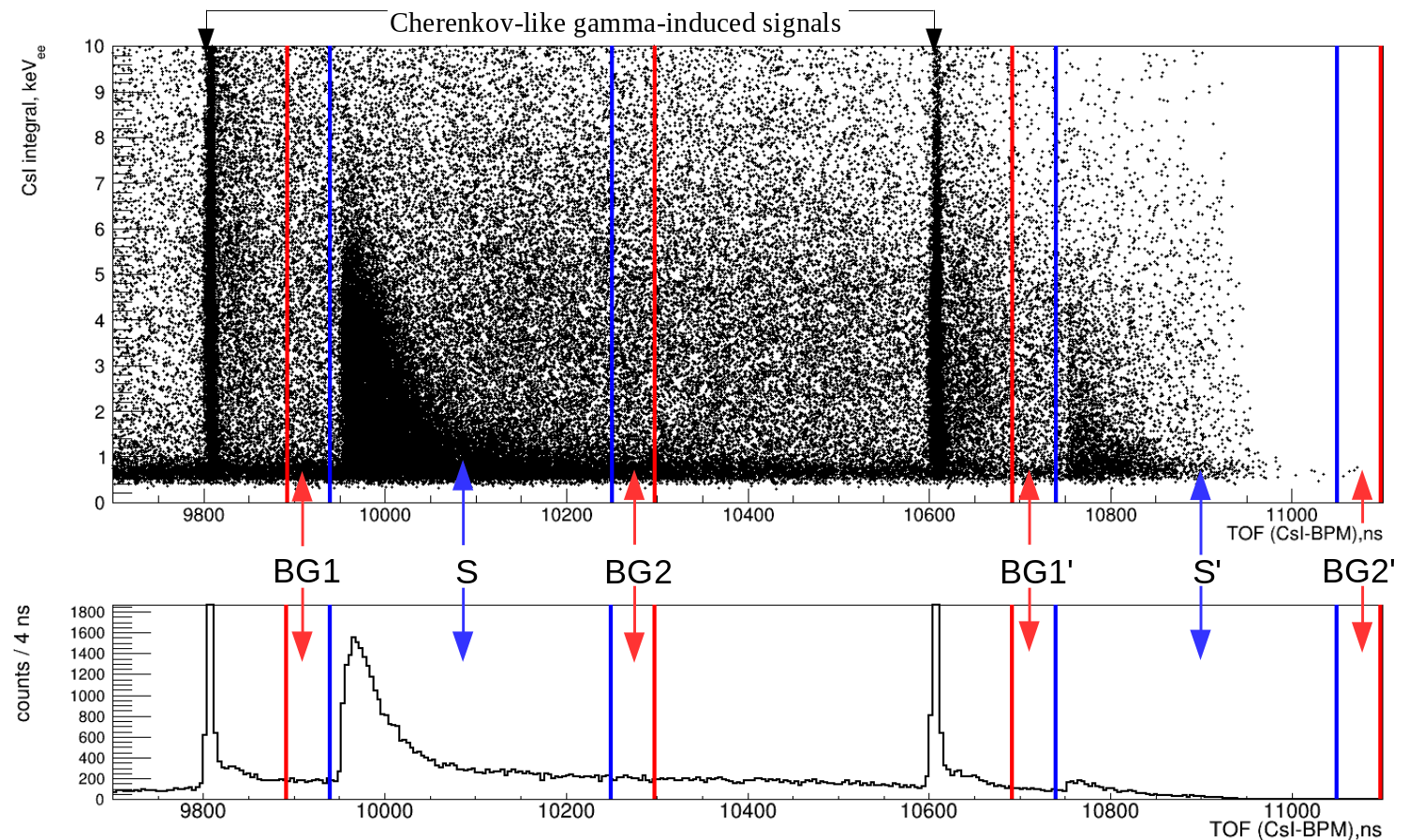}
\caption{\label{fig:coh4_tof} CsI[Na] energy deposition vs. time delay between its onset and BPM signal (top) for 1.26~MeV neutrons. Projection of the distribution onto the TOF axis for energies [1,10] keV$_{ee}$ (bottom). The signal (S) n-induced TOF region is marked with blue lines while background sidebands (BG) are 48 ns periods before and after each signal region. The gamma and neutron-induced populations marked with a prime are due to dependence between the CsI[Na] signal shape and its onset, connected to the trapezoidal trigger filter response (see text).}
\end{figure}

The CsI[Na] signals related to beam neutrons can be selected by the time delay between the BPM signal and CsI[Na] signal onset, as seen in Figure \ref{fig:coh4_tof}. It shows a narrow and energetic gamma-related TOF region followed by wider neutron TOF regions. The ``reflections'' of the gamma and neutron populations (marked with a prime in the figure) are due to dependence between the CsI[Na] signal shape and its onset, connected to the trapezoidal trigger filter response. Both channels of the ADC were triggered by the 50\% falling edge of the trapezoidal filter response to CsI[Na] signal. The timing of this 50\% falling edge is fixed within the waveforms. The delay between the 50\% falling edge point and a true onset of the CsI[Na] signal fluctuates from trigger to trigger. These fluctuations happen due to the variation in the arrival time of PMT pulses within the CsI[Na] scintillation decay times and affect the CsI[Na] signal shape. This variation is most pronounced at low CsI[Na] signal integrals and can be of the order of the BPM period --- 800 ns. This means that depending on the delay between the true onset of the CsI[Na] signal and the ADC trigger, the BPM waveform may contain a signal from one of two subsequent BPM periods. Such an ambiguity creates the aforementioned reflections in Figure \ref{fig:coh4_tof}. It deteriorates with the increase in the size of a CsI[Na] signal. We additionally suppress Cherenkov-like backgrounds in CsI[Na] by requiring the ratio of the integral in the first 100 ns to the full 3~$\mu$s integral to be lower than 0.5 and accepting only signals that have more than 5 PMT pulses. The cumulative efficiency of these cuts reaches $\sim$100\% at 11 PE corresponding approximately to 1~keV$_{ee}$. The background contribution is evaluated based on the background sideband windows, preceding and following the NR population in TOF. These sidebands are not expected to have beam-related signals in the integral region of interest (above 11 PE).

\begin{figure}[ht]
\centering
\includegraphics[width=1.0\textwidth]{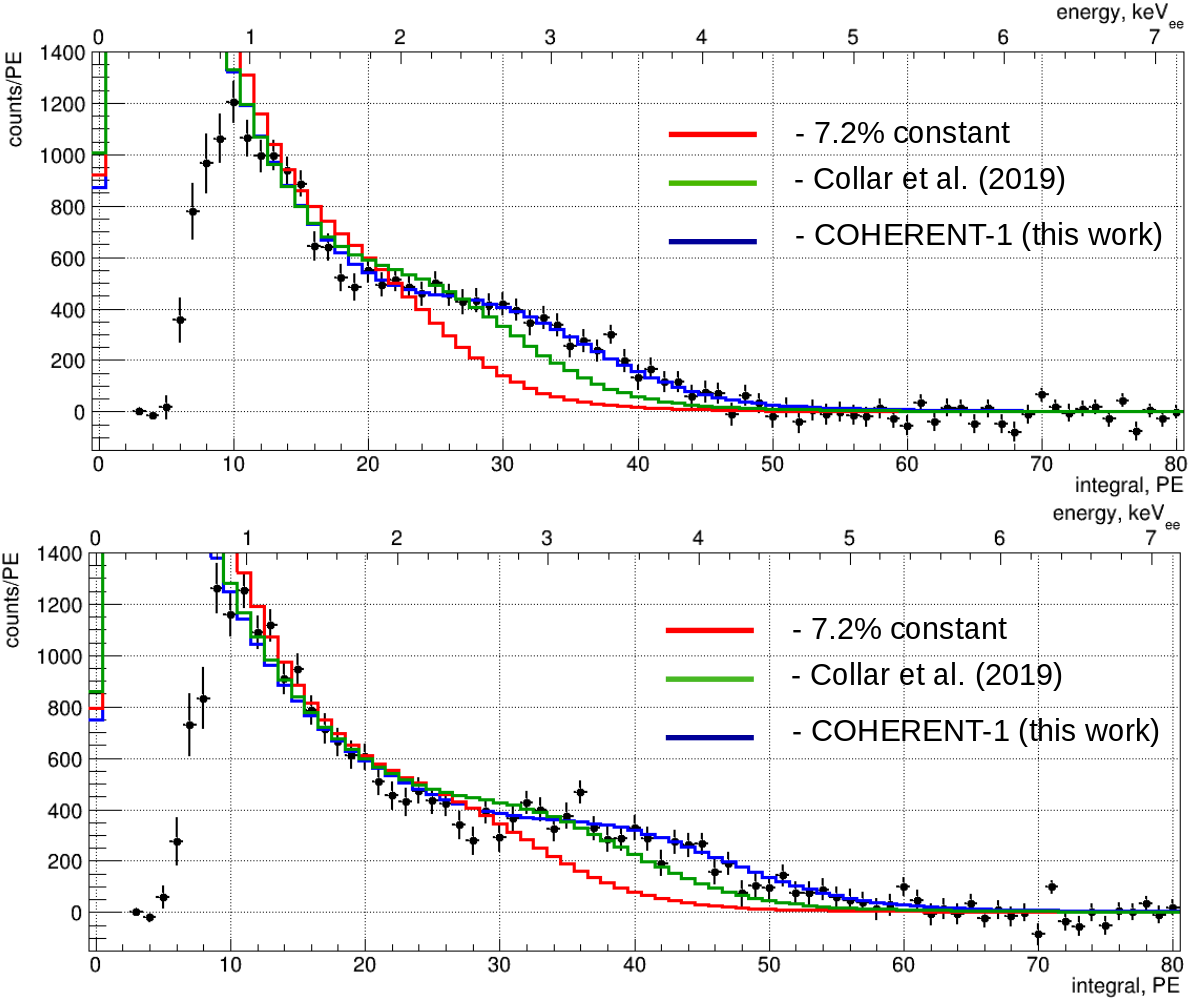}
\caption{\label{fig:coh4_nr} The integral spectrum from the COHERENT-4 measurements, top: 0.94 MeV neutron beam energy, bottom: 1.26 MeV; best-fit background model subtracted. Lines represent best fits for the models based on the existing QF value trends. The selection and trigger efficiency is very nearly 100$\%$ at 11~PE, but large effects are seen at lower energies.}
\end{figure}

We fit the observed NR spectra using a prediction based on a MCNPX-Polimi simulation. Three QF models were tested: a constant of 7.2\% suggested by the original analysis of COHERENT-2 \cite{Scholz_2017} for 6-16 keV NR, the best-fit model from  ref.~\cite{Collar_2019} and the trend from the updated COHERENT-1 results (Table \ref{Table_COH1}). The latter two were approximated by polynomial fits to data. We apply the quenching model taken at the respective energy to each neutron scattering act from the simulation. Then we sum quenched energy deposited by the same neutron and use this sum to produce a NR spectrum prediction. This approach allows us to account appropriately for the multiple scattering of neutrons. We complete the signal prediction with the background contribution determined from out-of-time data. The background rate for each sideband was fixed by the lengths of the timing windows. We observed differences in the background rate and spectra in the sidebands preceding (BG1) and following (BG2) the signal region. To account for this difference we take a weighted sum of the two in our signal region: $r*BG1+(1-r)*BG2$. This $r$ is profiled while fitting and allows the model to represent the rate of background in between two sidebands in the signal region. The fits are performed for the parts of the spectra with integrals larger than 11 PE. The trigger and selection efficiency is close to 100\% at such an energy deposition scale. The result of the fit for the 0.94 MeV neutron beam energy is shown in Figure \ref{fig:coh4_nr} (top). For this beam energy the fit value of $r$ for all QF models is very nearly 1.0 (fully BG1-like) so we can plot all the models after subtracting the same background spectrum. It can be seen that the COHERENT-1 model best fits the data. The $\chi^2/d.o.f.$ values for the constant 7.2\% model, the model from ref.~\cite{Collar_2019} and COHERENT-1 are 538/65, 180/65 and 60/65. When fitting the 1.26~MeV neutron data, the preferred $r$ value for the constant 7.2\% and ref.~\cite{Collar_2019} based QF models is still nearly 1.0 but lower for the COHERENT-1 model. For illustration, we overlay models on the residual spectrum using $r=1$ in Figure \ref{fig:coh4_nr} (bottom) making the discrepancy between data and model slightly worse for COHERENT-1 (2017/2020) compared to the best fit.  The COHERENT-1 model still achieves the best $\chi^2/d.o.f.$ of 82/65, comparison to 139/65 and 372/65 for the one from \cite{Collar_2019} and the 7.2\% model.

The COHERENT-4 measurement verifies the results of the tagged neutron COHERENT QF measurements presented in this work. The measurement is in very strong tension with both the constant $\sim$7.2\% QF trend suggested by \cite{Scholz_2017} and the model fit in \cite{Collar_2019}.

\section{Global CsI[Na] QF data fit} \label{sec:global_fit}

In this section we present a fit of available CsI[Na] QF data. We articulate the goals of the fit, discuss existing data, describe the fitting approach and quantify the uncertainties of the fit results.

\textbf{\underline{Goal.}} We perform a fit to allow accurate 
analysis of predictions in CEvNS measurements and studies of sub-GeV dark matter candidates using this scintillator at $\pi$ decay-at-rest neutrino sources. While the region of interest of NR energy for the former is from 5 to 30~keV$_{nr}$, the latter require a consistent description of the QF up to 60~keV$_{nr}$. We take care to understand the uncertainty of QF shape by performing a principal component analysis of the fit results.

\begin{figure}[htbp]
\centering
\includegraphics[width=0.6\textwidth]{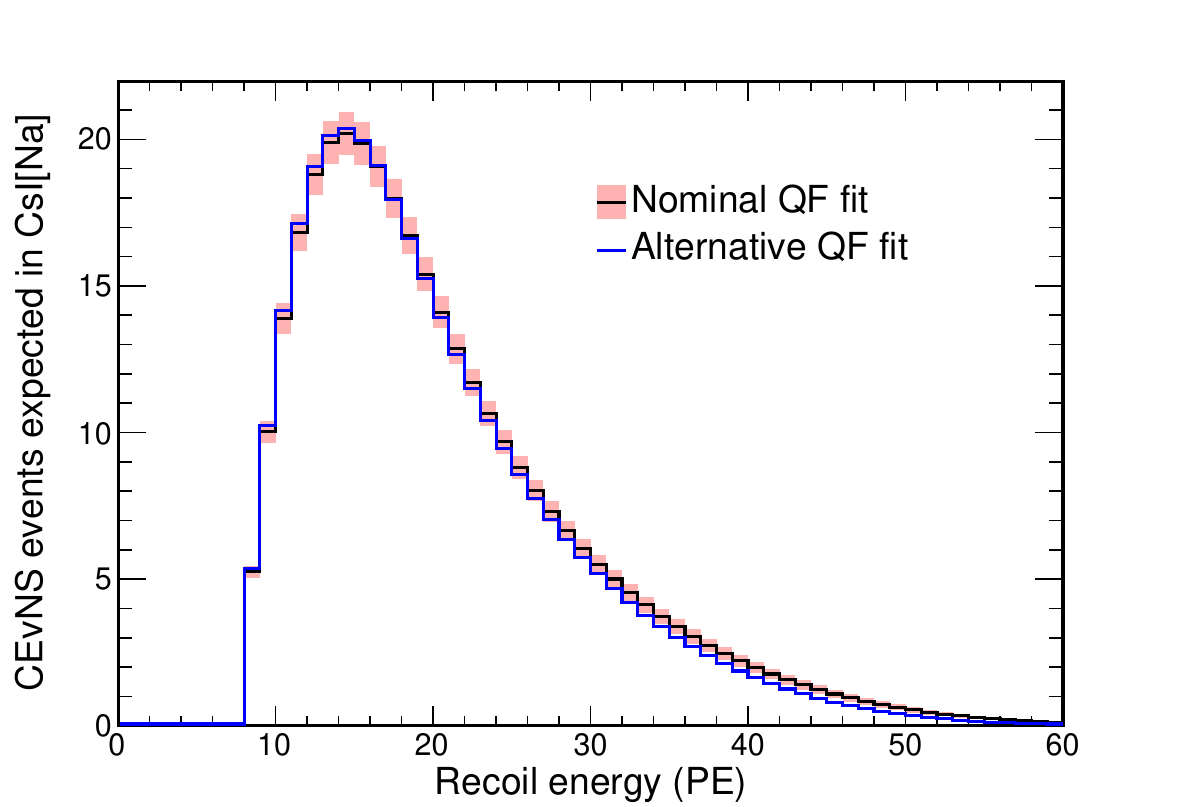}
\caption{\label{fig:QFfit_ucty} The recoil energy spectrum of expected CEvNS events in CsI[Na] described in \cite{CsI_CEvNS}. For the bulk of the distribution, the two fits agree quite well.  The total integrals of the two distributions differ by 2.6$\%$.}
\end{figure}

\textbf{\underline{Data.}} The existing measurements of the nuclear recoil quenching factor in CsI[Na] include ref.~\cite {Park_2002, Guo_2016, Collar_2019} and this work. Our fit includes the results of COHERENT-1/2/3 presented here as well as the Chicago-1 and Chicago-3 results from ref. \cite{Collar_2019}. All these datasets were collected with the same CsI[Na] crystal grown by the same manufacturer \cite{Proteus} using the same growing method and dopant concentration as the 14.6 kg crystal deployed at SNS.  We stress that the highest and lowest data points among these datasets are 3~keV$_{nr}$ and 63~keV$_{nr}$ which covers the entire NR energy range of interest. COHERENT-4, the endpoint measurement, is not included because the fit was used as a hypothesis test over a distribution of recoil energies rather than for measurements of the QF at specific energies. We do not include the previous data from refs.~\cite{Park_2002, Guo_2016}. The crystals used in these measurements have the dopant concentration in the range from 0.019 to 0.026~mole\% vs. 0.104~mole\% in ref.~\cite{Collar_2019} and COHERENT measurements. It is not clear whether the results of ref.~\cite{Park_2002} suggesting no dependence of QF on dopant concentration for 0.019--0.026~mole\% can be extrapolated up to 0.104~mole\%. Recent studies \cite{Collar_2019, Collar_2021} suggest no QF difference between a crystal with 0.104~mole\% and a cryogenic undoped crystal. Interpretation of such results is complicated by the difference between scintillation mechanisms of CsI[Na] and undoped cryogenic CsI and the contrast of results from ref.~\cite{Collar_2019} with the QF for alpha particles in the same material from ref.~\cite{Clark_2018}. Another difference between ref.~\cite{Park_2002, Guo_2016} and measurements included in the fit is signal integration time. COHERENT and ref.~\cite{Collar_2019} used 3~$\mu s$ while ref.~\cite{Park_2002}, apparently, used 7~$\mu s$ and ref.~\cite{Guo_2022} used 1.1~$\mu s$~\cite{Guo_2022}. Such a difference can lead to a systematic change in the QF if scintillation decay profiles for low energy nuclear recoils and 59.5~keV gamma rays from Am-241 are not the same. Our study suggests that in the $E_{nr}$ range from 20 to 60 keV QF(1.1~$\mu s$)/QF(3~$\mu s$)$\approx$1.08 and QF(7~$\mu s$)/QF(3~$\mu s$)$\approx$0.95 --- non-negligible systematic change.\footnotemark  If such a scaling is taken into account, the data from ref.~\cite{Park_2002} exhibit better agreement with results of ref.~\cite{Collar_2019} and this work, while the tension between data from ref.~\cite{Guo_2016} and other datasets increases. We have also performed a check of potential contributions from inelastic neutron scattering with gamma escape to results from ref.~\cite{Park_2002, Guo_2016}. The MCNPX-Polimi simulations incorporated neutron beam energy, geometry of setups and backing detector energy deposition requirements from ref.~\cite{Park_2002, Guo_2016}. The results of simulations suggest 5-10\% and 4-7\% lower $E_{nr}$ (and larger QF) for nominal 52 keV and 66 keV data points from ref.~\cite{Park_2002} depending on the proton quenching of the BC501a scintillator (see ref.~\cite{Verbinski_1968,Anghinofli_1979,Arneodo_1998,Kornilov_2009}). The contributions to other data points of ref.~\cite{Park_2002, Guo_2016} are negligible (few percent). The effect that data from ref.~\cite{Park_2002, Guo_2016} have on our fit, if included, is estimated with a sensitivity test (see ``Fit results and uncertainties'' and Fig.\ref{fig:QFfit_ucty}).

\footnotetext{We observe $QF(1\mu s)/QF(3\mu s) \approx QF(7 \mu s)/QF(3 \mu s)\approx1.0$ at 10~keV$_{nr}$. The increase of QF(1.0~$\mu s$)/QF(3~$\mu s$) up to $\sim1.08$ at 20~keV$_{nr}$ is confirmed by COHERENT-1/2/3. The decrease of QF(7.0~$\mu s$)/QF(3~$\mu s$) to $\sim0.95$ at 20~keV$_{nr}$ is consistent in COHERENT-1/3 (the  readout window of COHERENT-2 does not allow 7.0~$\mu s$ integration).}

\noindent\textbf{\underline{Fitting approach.}} We do not fit the QF curve directly due to correlated errors between the reconstructed visible energy and nuclear recoil energies in the ratio defining the QF. Rather, we calculate the scintillation response function which gives the visible energy expected for a given nuclear recoil energy. This allows a straightforward treatment of the errors involved, and is the physical construct needed for calculating signal spectra for CEvNS experiments. We empirically model the QF function as a fourth degree polynomial for 2-70 keV$_{nr}$. This polynomial is regularized by constraining the fit function to go through the origin. The use of a polynomial fit is a standard practice in a case of absence of a reliable physical model and tension between existing experimental data. We point out that a polynomial can describe quite well the modified Birks model from ref.\cite{Collar_2019} in CEvNS ROI. Data restricted to the CEvNS ROI are well-described by a lower order polynomial. The polynomial fits of the power from 1 to 3 of the data in the CEvNS ROI are contained, for the most part, in the uncertainty of our main fit. However, these fits fail to replicate trends in the observed data at higher recoil energies which are needed to predict other physical phenomena such as dark matter interactions. We must mention that the fit we perform utilizes the $E_{nr}$ uncertainties articulated in this work. The errors on the $E_{nr}$ are not given in ref.~\cite{Collar_2019}, and we thus consider the ``spread" quoted in that result as the $E_{nr}$ uncertainty estimate. A smaller estimate may be derived if the 3\% difference between measured and calculated neutron beam energy is considered to represent the whole $E_{nr}$ uncertainty.

\begin{figure}[htbp]
\centering
\includegraphics[width=0.49\textwidth]{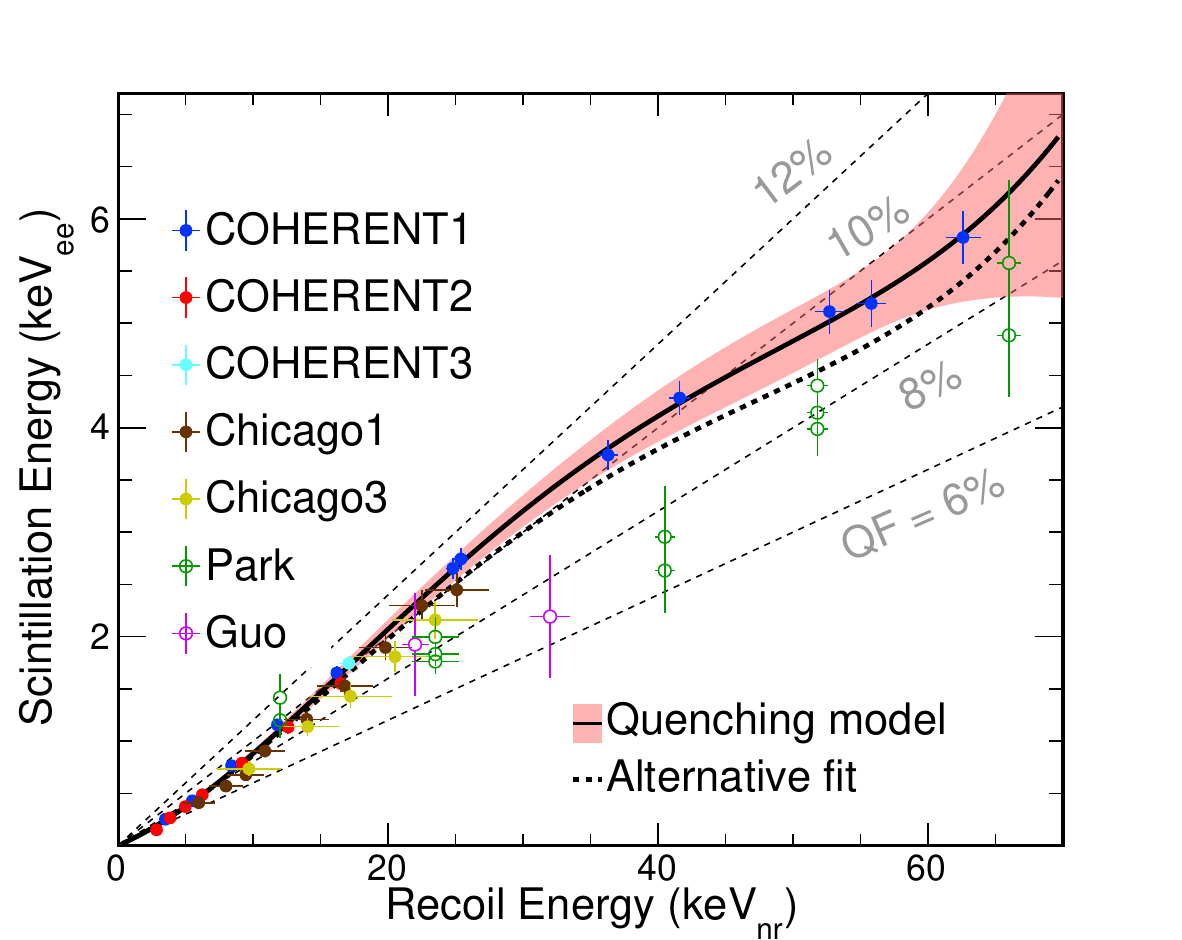}
\includegraphics[width=0.49\textwidth]{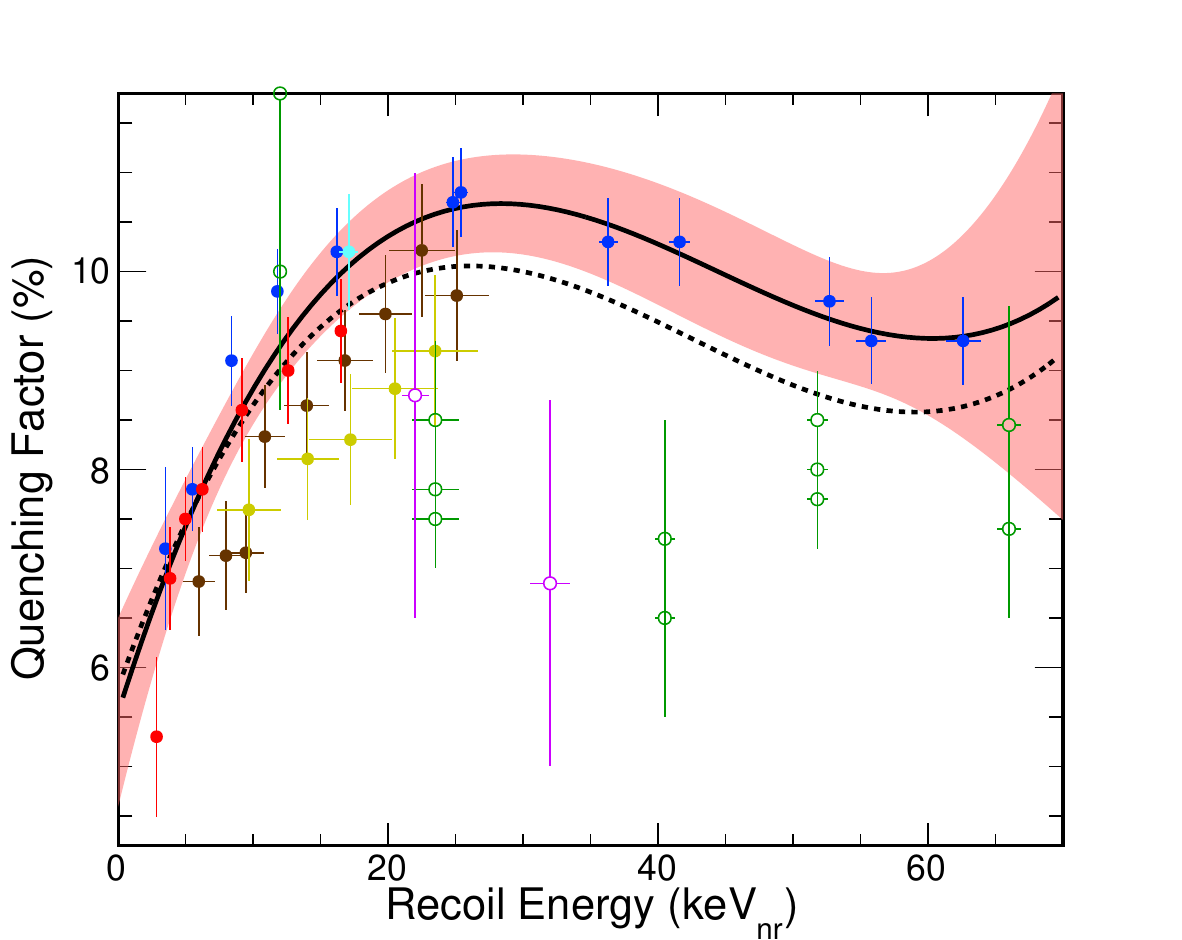}
\caption{\label{fig:QFfit} Quenching data in CsI[Na] along with our nominal best fit and error band. We fit the scintillation response curve (left) and also plot the QF (right). When plotting the quenching factor, the relative error in the recoil energy is propagated into the QF error. Data with empty circular points are not included in the nominal fit. The bold dashed line represents an alternative fit including data from \cite{Park_2002, Guo_2016} and a 3$\%$ uncertainty on $E_{nr}$ for data presented in~\cite{Collar_2019} (see text).}
\end{figure}

\noindent\textbf{\underline{Fit results and uncertainties.}} The best-fit function obtained with our approach is
\begin{equation}
    E_{vis}(E_{nr})=0.05546\times E_{nr} + 4.307\times E_{nr}^{2} - 111.7\times E_{nr}^3 + 840.4\times E_{nr}^4
\end{equation}
where $E_{vis}$ gives the visible detector response in MeV$_{ee}$ and $E_{nr}$ is the recoil energy in MeV$_{nr}$. We performed a principal component analysis to determine the error band, which will be included with the data release (see Section \ref{sec:share}). The $\chi^2/d.o.f.$ of this fit is 25.6/30. Due to the ambiguity of the NR energy uncertainty given in \cite{Collar_2019}, we perform a second fit where there is no error on the recoil energies for the Chicago-1 and Chicago-3 measurements. This yields a scintillation response curve within our fit error band, but the $\chi^2/d.o.f=47.4/30$ is significantly larger than $1$. We thus calculate a reduced $\chi^2_{red}$ defined so that $\chi^{2}_{red}/d.o.f.\leq1$ for both fits. Our error band is drawn according to this $\chi^2_{red}$ which, consequently, increases the QF error band size by 58$\%$. This gives our full quenching model, shown in Figure~\ref{fig:QFfit}. We point out that such an expansion of the uncertainty in case of disagreement between measurements is in accord with the PDG approach~\cite{PDG_fit}. The total uncertainty gives a $\sim$~4$\%$ error in the expected CEvNS rate for a $\pi$ decay-at-rest flux~\cite{CsI_CEvNS}, much smaller than the 25$\%$ used in \cite{Akimov_2017}. We perform two additional cross-check fits (not shown in Figure~\ref{fig:QFfit}) which also yield curves within our error band: a fit including data from \cite{Park_2002, Guo_2016} and a fit restricted to the CEvNS region of interest ($5 \leq E_{nr} \leq 30$~keV$_{nr}$).

To further validate the assumptions in this fit, we show the CEvNS prediction for a $\pi^+$ decay-at-rest neutrino flux given our quenching fit and uncertainties in Figure~\ref{fig:QFfit_ucty}.  We also show an alternative fit including data from \cite{Park_2002, Guo_2016} and a 3$\%$ uncertainty on $E_{nr}$ for data presented in ref.~\cite{Collar_2019} (bold dashed line in Figure~\ref{fig:QFfit}). The expression for such a fit is

\begin{equation}
    E_{vis}(E_{nr})=0.05545\times E_{nr} + 4.297\times E_{nr}^{2} - 111.3\times E_{nr}^3 + 836.4\times E_{nr}^4
\end{equation}
where $E_{vis}$ gives the visible detector response in MeV$_{ee}$ and $E_{nr}$ is the recoil energy in MeV$_{nr}$. The predictions assuming the nominal and alternative quenching fits agree in the peak below 25~PE (roughly 20~keV$_{nr}$), see Figure~\ref{fig:QFfit_ucty}. However, the two fits are discrepant at about 1~$\sigma$ beyond in the tail of the distribution beyond 35~PE (roughly 27~keV$_{nr}$). The total CEvNS event rate predicted by the two fits agree to 2.6$\%$ (341 counts for the default fit vs. 332 for the alternative fit in assumptions from ref.~\cite{CsI_CEvNS}), lying within the predicted uncertainty of our quenching model.  Dark matter produced at the SNS would produce more energetic recoils with a substantial fraction of events expected between 40 and 60~PE for some dark matter mass assumptions. Though the two quenching models are more discrepant in this region, the selection efficiency is nearly 1 for such high-energy recoils, and thus the high-recoil quenching does not strongly influence the dark matter interaction rate which limits the search.  

\section{Data sharing} \label{sec:share}

In order to increase transparency of our study we intend to share the raw data of COHERENT CsI[Na] QF measurements together with a short technical note providing discussion of certain analysis details and benchmark values. The data release is under preparation and will be published shortly after this work.

\section{Conclusion} \label{sec:concl}

In this work we presented results of CsI[Na] nuclear recoil quenching factor measurements performed by the COHERENT collaboration and aimed to reduce the systematic uncertainty of the COH-CsI CEvNS measurement at the SNS. We obtained quenching factor values mostly larger than in previous experiments \cite{Park_2002,Guo_2016,Collar_2019} and performed a fit of global data, thereby reducing the former CEvNS rate uncertainty from a leading 25\% to a modest 4\% in the analysis of the full COH-CsI dataset, subdominant to the SNS neutrino flux uncertainty of 10\%. Scintillating CsI[Na] crystals have long been recognized for their potential to make a significant impact on CEvNS measurements. These new QF measurements and careful reexamination of previous data contribute to better understanding of CsI[Na] response for current and next-generation CEvNS experiments.

\section{Acknowledgements} \label{sec:aknow}

The COHERENT collaboration acknowledges the resources generously provided by the Spallation Neutron Source, a DOE Office of Science User Facility operated by the Oak Ridge National Laboratory. The Triangle Universities Nuclear Laboratory is supported by the U.S. Department of Energy under grant DE-FG02-97ER41033. This work was supported by the Ministry of Science and Higher Education of the Russian Federation (Project ``Fundamental properties of elementary particles and cosmology'' No. 0723-2020-0041); the Russian Foundation for Basic Research (proj. \# 17-02-01077 A); the DOE HEP grant DE-SC0020518; the National Research Foundation of Korea (NRF 2022R1A3B1078756). This research used the Oak Ridge Leadership Computing Facility, which is a DOE Office of Science User Facility. We are thankful to J.~Collar, B.~Scholz and Z.~Wang for useful exchanges.

\appendix

\section{Notes on methodology} \label{sec:meth}

In this section we describe essential aspects of the CsI[Na] QF analysis, which may differ in the literature \cite{Guo_2016,Park_2002,Collar_2019}, and provide information relevant to understanding and interpretation of our results.

\subsection{Signal integration and SPE pulse shape} \label{subsec:spe_int}

\begin{figure}[htbp]
\centering
\includegraphics[width=.7\textwidth]{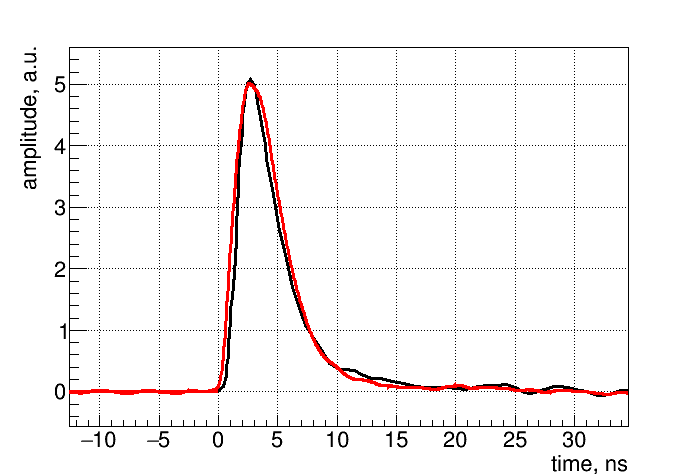}
\caption{\label{fig:spe_shape} Red: H11934-200 signal shape acquired in the calibration with 30 ps laser pulses. Black: digitized time response plot of a H11934 PMT from the manufacturer \cite{H11934} (70 ps light pulse)}.
\end{figure}

The response of CsI[Na] to nuclear recoils is evaluated based on integration of signals measured as ADC waveforms. With the exception of the original COHERENT-1 result \cite{Grayson_2017}, all our QF measurements use pulse finding to integrate waveform signals. This improves resolution by taking into account only signal-related samples rather than a whole integration window (3~$\mu$s) and reduces potential bias from shifts in the DC baseline amplitude. The pulse endpoints are identified by a threshold ADC value. For low energy signals, the choice of threshold can have a significant impact on results. Low thresholds are more susceptible to overintegration by accepting noise fluctuations of the baseline, while high thresholds can underintegrate by missing portions of the true PE pulse. We verified our pulse finding and integration approach with a waveform simulation. Such a simulation incorporated the noise RMS determined by data and an average SPE shape from the dedicated PMT calibration. During that calibration, the H119340-200 unit was illuminated by 30 ps laser pulses, each generating about a 10 PE response in the PMT. With a robust electron transition time spread of 0.27 ns \cite{H11934}, the average pulse shape of these signals should be very close to the SPE pulse shape. Averaging of individual signals was performed using a LECROY WaveRunner 640Zi, 20GS/s digital oscilloscope. The PMT time response obtained in our measurement (see Figure \ref{fig:spe_shape}) is close to that provided by the manufacturer, though wider by about 1 ns. Importantly, both shapes demonstrate a relatively long low-amplitude tail which can be lost in noise. The shapes suggest that 95\% of SPE integral are contained within 11 ns of the pulse peak and 98\% within 17 ns. The remaining 2\% is spread up to 30 ns away from the peak. The results of our waveform simulation tests show a 1-3\% bias in reconstructed integral for COHERENT datasets which is included in the total uncertainty budget.

\subsection{Determining background model} \label{subsec:bg_choice}

QF experiments using a tagged nuclear recoil approach benefit from the reduction of accidental background contamination in the response spectra. The remaining background contribution can be taken into account with the help of representative background subselections. We define two time intervals to characterize them. The first one is ``pre-beam''; it precedes the arrival of neutrons to the scatterer. The second is ``in-beam''; it includes the arrival of neutrons and covers delayed onsets of CsI[Na] response to NR. Two approaches to determine background subselections or models may be called pre-beam-based and in-beam-based respectively. The key features of these approaches are listed in Table \ref{table:Table_BG}. 

The first approach utilizes the pre-beam events coincident with the $n$-like BD signals. It coincides with the signal NR selection in everything but the time interval choice. The rate of background events from the pre-beam interval can be matched to that from the in-beam signal ROI by simple scaling based on the time interval lengths. After such a scaling the background contribution can be subtracted from the signal ROI spectrum. The residual represents the NR-related integral distribution of interest. Despite its attractive simplicity the pre-beam based approach has drawbacks. First, the number of events in the pre-beam background spectrum is of the same order of magnitude as in the signal spectrum. Subtraction of spectra increases the statistical uncertainty of the residual, decreasing the precision of the QF value estimates. Second, the estimate of the background contribution to the signal ROI evaluated from the time interval lengths is biased and requires correction. In the absence of beam-related signals the time distribution of observed onsets is driven by the appearance of random afterglow pulses. The distribution corresponds to the Poissonian probability to observe $k=0$ pulses within time $t$ from the start of the ROI:

\begin{equation}
  \label{eq:pois1}
  P(k=0, \lambda=\mu \cdot t)=e^{-\mu \cdot t},
\end{equation}
where $\mu$ is the probability for an afterglow pulse to appear per unit time. Indeed, the distribution seen in Figure \ref{fig:coh2_tof} (right, bottom panel) is monotonically decreasing. It is not straightforward to understand by looking at only a signal-rich subselection like one in Figure  \ref{fig:coh2_tof} (left, bottom panel). Finally, the integral distribution of a background subselection may be distorted. If an accidental afterglow PE arrives shortly before an actual NR signal, the onset of such an event is misidentified as pre-beam. The background subselection in this case turns out to be affected by NR signals. The pre-beam signals with 10-50 PE in Figure \ref{fig:coh2_tof} (left, top panel) are of this nature. The bias induced by this pathology depends on the afterglow intensity (see also Appendix~\ref{subsec:afterglow}). Requiring a coincidence of several PE to identify an onset would eliminate this problem, but would also affect acceptance of low energy signals.

\begin{table}
\begin{center}
\parbox{1.0\linewidth}{\caption{\label{table:Table_BG} Determining CsI[Na] background subselections.}}
\begin{tabular}{p{1.5cm} c c p{3.5cm} p{4.2cm}}
\hline
\centering Approach & BD signal & Time interval &\centering  BG rate scaling & \centering Illustration (Figure \ref{fig:coh2_tof}) \tabularnewline \hline
\centering Pre-beam based & $n$-like  &\centering  pre-beam & fixed by in-beam to pre-beam length ratio &\centering left, -1000~ns to -100~ns \tabularnewline
\centering In-beam based  & $\gamma$-like & in-beam &\centering  fit to the NR spectrum &\centering right, -100~ns to 1000~ns
\end{tabular}
\end{center}
\end{table}

The in-beam-based approach relies on selection of events associated with the gamma-induced signals from the BD in coincidence with the beam. The number of BD signals induced by environmental gamma rays is much larger than that from neutrons (see Figure \ref{fig:coh2_psd}, Figure \ref{fig:coh3_psd}). Thus, one obtains a greater statistical power of the background subselection --- its spectrum can be used as a PDF with negligible shape uncertainty. The drawback of using gamma-induced triggers is that some of them are beam-related. Those can be excluded based on the TOF calculated with BD and BPM (Figure \ref{fig:coh3_tof}). Or, it can be demonstrated that there is no significant excess with beam-like TOF based on CsI[Na] and BD signals onset (Figure \ref{fig:coh2_tof}, right panel). We utilized the in-beam based approach to determine the background model for COHERENT-1/2/3 and a variant of the pre-beam based approach for the COHERENT-4 measurement.

\subsection{Effect of CsI[Na] afterglow on QF data analysis} \label{subsec:afterglow}

Crystals of CsI[Na] demonstrate continuous phosphorescence following a large energy deposition on time scales up to a millisecond \cite {Collar_2015}. It may happen that accidental afterglow pulses appear in a 3 $\mu$s integration window of NR signals, biasing the QF estimate to larger values. Unfavourable influence of the afterglow can be suppressed by requiring fewer than N pulses in the CsI[Na] waveform pretrace~(PT)~--- a time region preceding the arrival of beam neutrons. Table~\ref{table:Table_AG} shows cuts on the pretrace occupancy we used in the analysis of COHERENT data. It also specifies corresponding afterglow contamination values. The latter were evaluated in situ by measuring the count rate in out-of-time data after imposing the same restriction on PT occupancy. In the COHERENT-2/3 fits the afterglow contamination is accounted for by convolving the corresponding integral distribution with the integral spectrum prediction. For COHERENT-1 this effect is included as an uncertainty. Afterglow contamination in COHERENT-4 data is negligible as the relevant signals are of much higher energy.

\begin{table}
\begin{center}
\parbox{1.0\linewidth}{\caption{\label{table:Table_AG} The afterglow contamination expected in a 3 $\mu$s integration interval for COHERENT data.}}
\begin{tabular}{c c p{1.5cm} c c c}
\hline
Dataset    & PT length, $\mu$s & \centering PT cut (~$\leq$~pulses) & Mean contamination, PE & NR signal scale, PE\\ \hline
COHERENT-1 &  5.0  & \centering 1 & 0.2 & 3--80  \\
COHERENT-2 &  1.0  & \centering 0 & 0.3 & 3--25 \\
COHERENT-3 &  10.6 & \centering 5 & 1.1 & $\sim$ 21 \\
COHERENT-4 &  7.0  & \centering 1 & 0.1 & 11--60
\end{tabular}
\end{center}
\end{table}

The misidentification of a signal onset and consequent distortion of the integral may also happen due to a spurious afterglow pulse. In our analyses, the time ROI within the digitized CsI[Na] traces is closely aligned with the expected neutron arrival time. The time interval before the first signal PE is quite small for an afterglow pulse to fit in ($\sim$100 ns). The chance of onset misidentification is just a few percent for all but the lowest-energy nuclear recoil signals of COHERENT-1/2 for which the onset can be delayed by up to 1 $\mu$s. For these low-energy recoils, the onset can be misidentified in $\sim$~10\% of events with $\sim$30\% distortion of the measured integral for each ($\pm$1 PE on a scale of a 3 PE signal). This gives a $\sim$3\% total bias to the nuclear recoil distribution which is negligible in comparison to the total uncertainty budget for these low energy NR (see Table \ref{Table_COH1} and Table \ref{Table_COH2}).

\subsection{QF definition at low NR energy} \label{subsec:def_lowE}

Due to the relatively slow CsI[Na] scintillation decay times \cite{Collar_2015}, the first observed PE from a NR can be delayed relative to the interaction time. The delay between the interaction time and the first PE arrival depends on the number of generated PE. We simulate this dependence using the empirically averaged time profile of the CsI[Na] response to 59.5 keV gamma rays. The mean delay time is expected to vary from 40~ns for 30~PE signals to 280~ns for 3~PE signals. It is negligible compared to the 3~$\mu$s integration window for larger signals, but can be significant for lower energies. We thus define the QF we measure as ``effective'', integrating all light in a 3~$\mu$s window starting from the first PE appearing after the expected neutron interaction time in CsI[Na]. In general, the QF as defined in this work is slightly larger than the ``real'' QF. This particular definition of QF was chosen to be consistent with the CEvNS data analysis of COH-CsI experiment \cite{Akimov_2017}.

\subsection{Inelastic interactions and QF measurements} \label{subsec:inel_contr}

The QF measurements based on the tagged neutron elastic scattering predict NR energy from kinematic considerations and known incident neutron energy. This strategy fails if a large fraction of neutron interactions are inelastic. Often such interactions are accompanied by emission of secondary particles which shift the observed energy out of the NR energy region of interest. However if the interaction only produces a relatively high energy gamma ray, the gamma can escape the crystal with no energy deposition. In this case, the expected NR energy is lower than would be expected for elastic scattering. The exact contribution of inelastic interactions depends on the strictness of TOF and BD integral cuts. We take it into account using a MCNPX-Polimi simulation, so does ref.~\cite{Collar_2019}.

\section{H11934-200 PMT response linearity} \label{sec:pmt}

The direct proportionality of the PMT anode output current and the incident light intensity is usually referred to as the PMT response linearity. At high illumination levels the PMT signal may deviate from the ideal linearity. A decreased signal-to-light ratio with respect to linear expectation is often referred to as the PMT ``saturation'', while an increased ratio may be referred to as ``overlinearity''. In accordance with ref. \cite{Hamamatsu_2007} there are two factors limiting the anode current linearity: the distortion of the interstage voltage of a divider circuit by high dynode/anode currents and space charge effects. While the former affects mostly the direct current operation mode of the PMT and depends on the average anode current value, the latter affects the pulsed operation mode and depends mostly on the peak signal current. Linearity of a PMT is determined by measuring the ratio between the PMT anode current and the incident light intensity. The relative scale of the latter may be established by collimating light to known apertures, using calibrated neutral density filters or by adjusting the intensity of the light source monitored by the reference PMT. The non-linearity manifests itself through deviations of a signal-to-light ratio from the constant obtained in low anode current operation mode. Descriptions of different PMT linearity tests and schematic drawings of the corresponding setups can be found in ref. \cite{Hamamatsu_2007}.

In ref.~\cite{Collar_2019} the results of COHERENT-1/2 were reconsidered based on a claim that the Hamamatsu H11934-200 PMT used to read out the CsI[Na] crystal exhibited charge saturation at the 59.5~keV ER calibration line. Such a saturation, if present, would have led to the bias of the measured QF to larger values. The claim was made based on a decreasing trend of the crystal light yield estimate vs. the PMT bias voltage. In this Appendix we present the results of the PMT characterization performed to scrutinize the claim of ref. \cite{Collar_2019}. In Section~\ref{subsec:scale} we describe the PMT response to the scintillation from 59.5 keV and compare it to the PMT specifications. In Section~\ref{subsec:all_tests_lin} we confirm the PMT linearity in the region of interest with the conventional tests. Finally, in Section~\ref{subsec:LY_and_SPE} we show that a light yield estimates in units of PE/keV at different PMT bias voltage can depend on the model of the SPE charge distribution and should not be used to characterize the linearity of a PMT.

\subsection{Scale of the 59.5 keV$_{ee}$ signal in CsI[Na]} \label{subsec:scale}

\begin{figure}[htbp]
\centering
\includegraphics[width=1.0\textwidth]{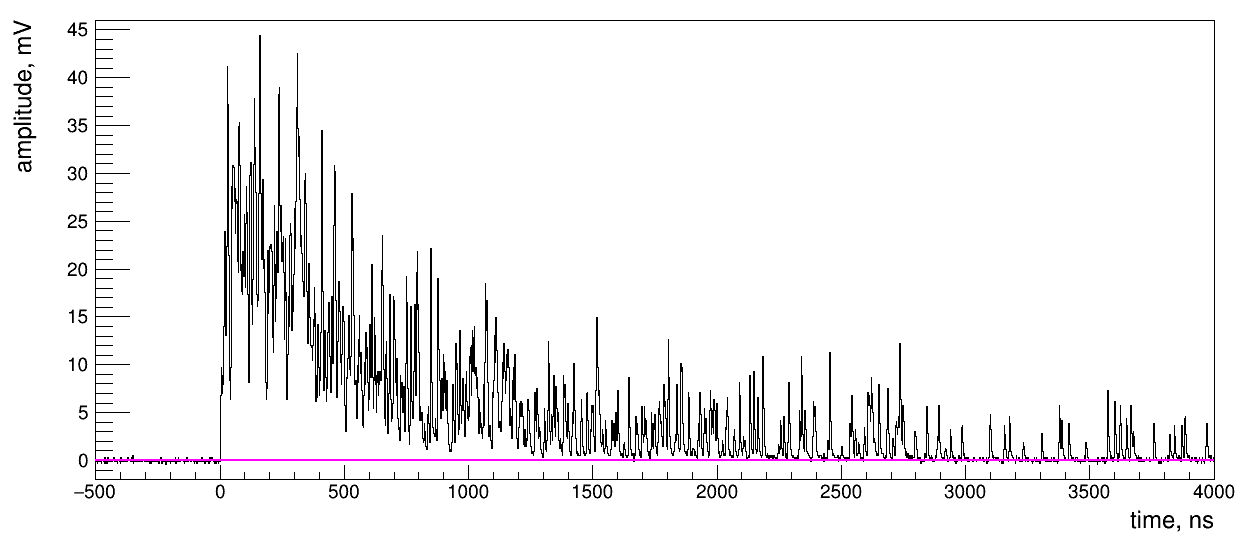}
\caption{\label{fig:coh_am_shape} Example of a 59.5 keV signal waveform in COHERENT-2 data.}
\end{figure}

The CsI[Na] crystal response to a 59.5 keV gamma ray generated about 1000 PE in the H11934-200 PMT at the time of COHERENT-1/2 measurements. A representative signal shape can be found in Figure \ref{fig:coh_am_shape}. Due to the relatively slow timescales of CsI[Na] scintillation \cite{Collar_2015}, such a signal has a maximum amplitude of about 15 times higher than the typical SPE amplitude and occupies several microseconds of a waveform. In the COHERENT-2 experiment this corresponded to an integral of $\sim$20 nVs and $\sim$40 mV maximum amplitude.

One can estimate the anode current induced by a 1000 PE signal of the PMT and compare it with the linearity limit published in the H11934-200 spreadsheet \cite{H11934} by the manufacturer. As a conservative approach, we assume 1000 PE are generated within 300 ns, much shorter than the 3~$\mu$s integration window. Biased to -950 V, the PMT operates at a gain of 4.7$\times$10$^6$, detecting up to 7.5$\times$10$^{-10}$~C of the anode charge from a 59.5~keV$_{ee}$ gamma ray.\footnotemark
\footnotetext{The H11934-200 datasheet \cite{H11934} suggests the $2\times10^6$ typical gain at -950 V bias voltage. In this estimate we use our own measurements. The gain curve can be described well with a power law ($9.84\pm0.05$) function in the range from -600 to -960~V. The extrapolated power law exceeds the data by about 6\% for -970 to -1000~V and up to 15\% at -1050~V.}
In our conservative approach, this gives a pulse anode current value of $\sim2.5$~mA, which is considerably lower than the published limits for linearity: 20 mA for $\pm2\%$ non-linearity and 60 mA for $\pm5\%$ (at the bias voltage of -900~V).\footnotemark Ref.~\cite{Collar_2019} suggests 13\% and 6\% bias due to non-linearity for the COHERENT-1 and COHERENT-2 measurements respectively which is inconsistent with the PMT specifications~\cite{H11934}.

\footnotetext{According to ref. \cite{Hamamatsu_2007} any dynode type provides better linearity when the bias voltage is increased. This means that comparing the linearity limit evaluated at -900 V to the anode current measured at -950 V is a conservative estimate.}

\subsection{PMT linearity tests} \label{subsec:all_tests_lin}

We conducted several tests both with the CsI[Na] crystal and controlled light sources to verify the linearity of the PMT. We stress that it is exactly the same PMT and base assembly that was used in the COHERENT-1/2/3/4 measurements while ref.~\cite{Collar_2019} used a different unit but of the same model. The CsI[Na] crystal used for linearity tests in our study was not the unit used in the COHERENT QF measurements, but it is of the same size and manufactured by the same company \cite{Proteus}.

\begin{figure}[htbp]
\centering
\includegraphics[width=1.0\textwidth]{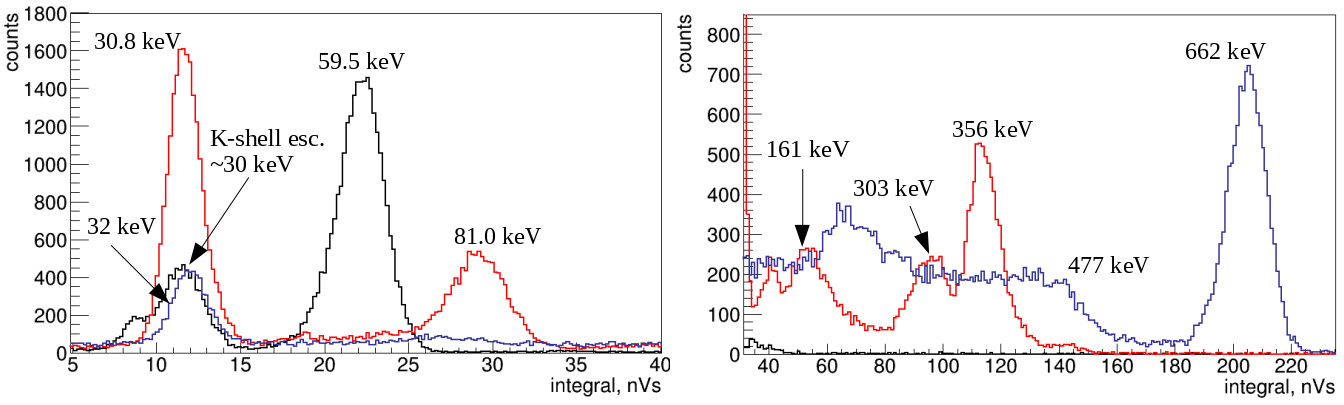}
\caption{\label{fig:peaks_ly} CsI[Na] response to $^{241}$Am (black), $^{133}$Ba (red) and $^{137}$Cs (blue) gamma calibration lines measured by the H11934-200 PMT when biased to -950~V. The spectrum is divided into 5-40 nVs (left) and 30-230 nVs (right) parts and resampled for clarity.}
\end{figure}

\begin{figure}[htbp]
\centering
\includegraphics[width=1.0\textwidth]{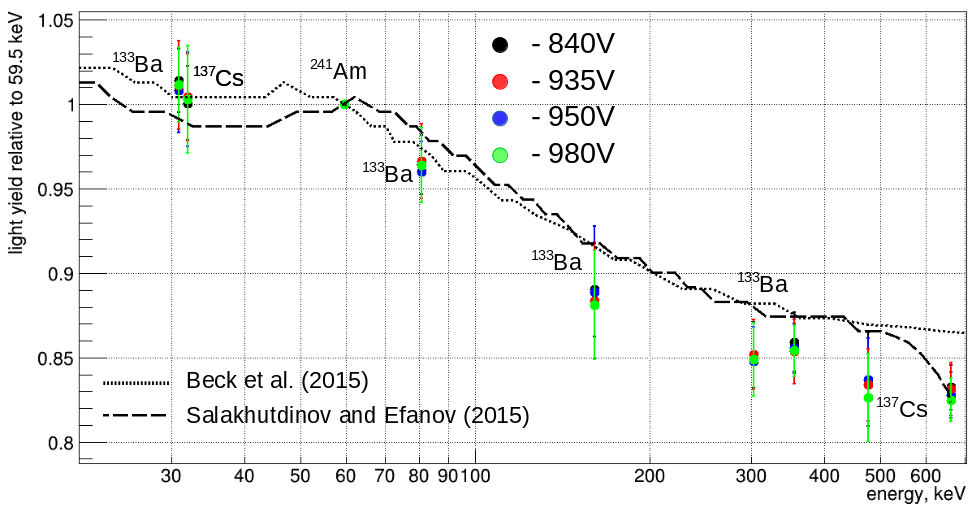}
\caption{\label{fig:rel_ly} Light yield of the CsI[Na] crystal relative to 59.5~keV for -840~V (black), -935~V (red), -950~V (blue) and -980~V (green) bias voltage and data driven models from refs. \cite{Beck_2015,Salakhutdinov_2015}. Though we observe non-linearity in the crystal response, the relative light yield is consistent for all bias voltages.}
\end{figure}

First, the relative light yield for a variety of calibration lines from $^{241}$Am, $^{133}$Ba and $^{137}$Cs sources was measured at PMT bias voltage values of -840, -935, -950 and -980~V. For each bias voltage value we normalize the CsI[Na] response by the one observed at 59.5 keV. The calibration lines from 30.8 to 662~keV span about 18 times~the PMT signal magnitude with additional factor of $\sim4$ from changes in the gain --- from 4 to 290 nVs.\footnotemark
\footnotetext{Taking into account the non-linearity of the CsI[Na] scintillation response (see also Figure \ref{fig:rel_ly}).}
We estimate the light yield of the CsI[Na] crystal to be 13~$\pm$~3~PE/keV at 59.5 keV with uncertainty dominated by a low precision of the SPE characteristic charge value related to the digitization rate, 8 ns/sample. We used the same 3~$\mu$s signal integration window defined for our QF and CEvNS studies. Figure \ref{fig:peaks_ly} shows the CsI[Na] response to the calibration lines from the sources used in the test. The relative light yield plot is shown in Figure \ref{fig:rel_ly}. It does not depend on the PMT bias voltage, which would be expected if the PMT response is significantly non-linear. Moreover, the observed values of relative light yield agree with the published data on the non-linearity of CsI[Na] crystal light response \cite{Aitken_1967,Mengesha_1998, Salakhutdinov_2015, Beck_2015} within our experimental uncertainty. The results of this test do not contradict the PMT specifications, but are in tension with the claim of ref.~\cite{Collar_2019}.

\begin{figure}[htbp]
\centering
\includegraphics[width=1.0\textwidth]{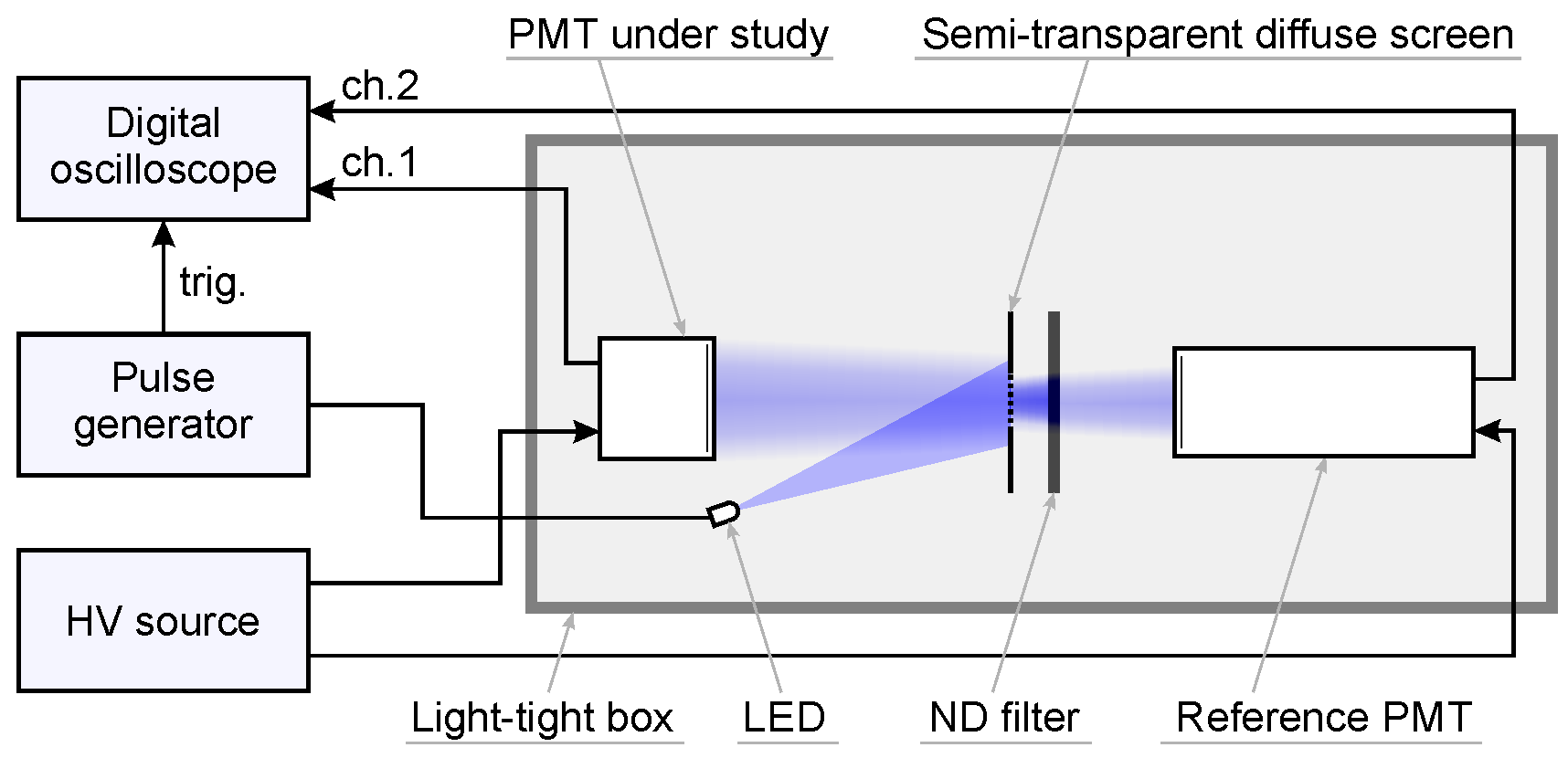}
\caption{\label{fig:Setup2} The experimental setup to measure the pulse charge linearity limit of the H11934-200 PMT at high light intensity.}
\end{figure}

\begin{figure}[htb]
\centering
\includegraphics[width=1.0\textwidth]{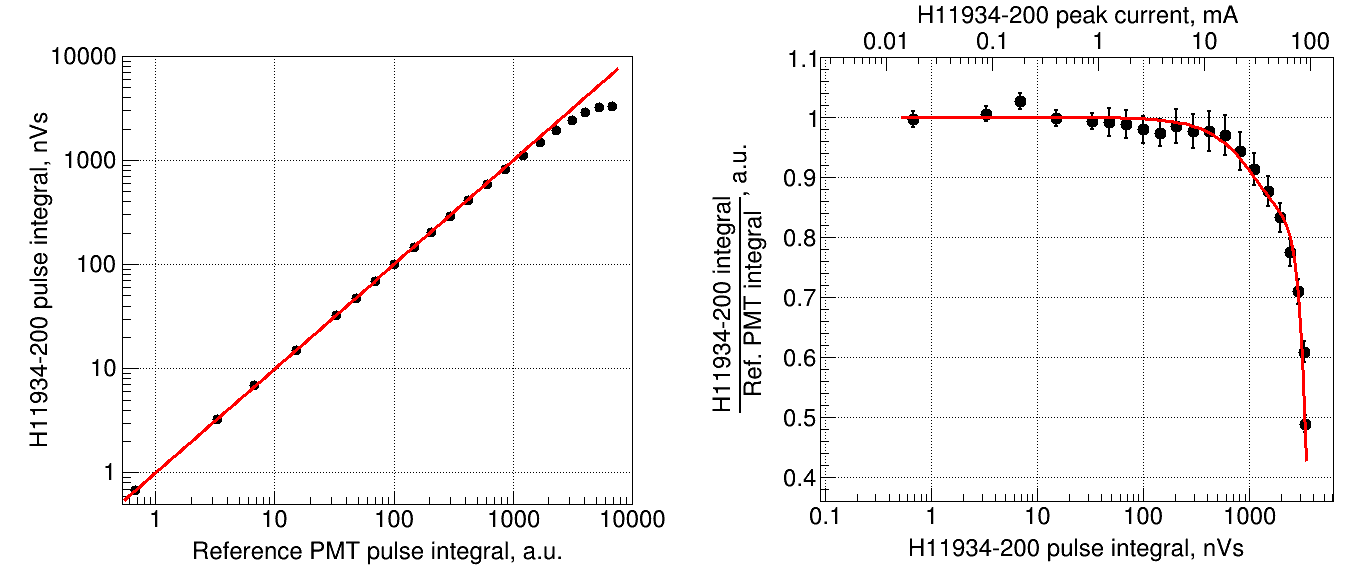}
\caption{\label{fig:nl_pulse} Left: H11934-200 pulse integral vs. reference PMT pulse integral, red line corresponds to direct proportionality. Right: ratio between the pulse integrals of H11934-200 and the reference PMT vs. the H11934-200 pulse integral. The red line shows the result of an ad hoc fit to the polynomial of the fourth degree. The uncertainties increase with the pulse integral due to the scaling coefficients correcting for the changes in the reference PMT bias voltage and the dynamic range of the digital oscilloscope.}.
\end{figure}

We continued characterization of the H11934-200 by measuring the pulse charge linearity limit using a 470~nm LED and a reference FEU-143 PMT. For this, an experimental set up shown in Figure~\ref{fig:Setup2} was used. The 750 ns long pulses of LED light were directed onto a semi-transparent diffuse screen, uniformly illuminating the PMT under study and also the reference PMT. Over the course of the measurement, the electric pulse amplitude biasing the LED was increased incrementally, thus increasing the illumination intensity and the signal pulse amplitude at the anodes of both PMTs.  The reference PMT stayed within the linear regime throughout data-taking.
The linearity of the reference FEU-143 PMT was achieved by switching to a lower bias voltage once the signal amplitude approached the saturation region at the current bias voltage. The saturation regions were determined in advance using a set of calibrated neutral density (ND) filters. The curve of the H11934-200 signal charge (at constant bias voltage of -950V) vs. the reference PMT signal charge explicitly tests the linearity of the former. To ensure no anode current saturation effects could influence the results, the repetition rate of the illumination was chosen as low as 50~Hz. The results are presented in Figure~\ref{fig:nl_pulse}. The region of $\pm3\%$ pulse linearity extends up to $\sim$400 nVs charge with a 750~ns pulse length. This scale corresponds to $\sim$10~mA peak anode current, which is of the same order of magnitude as the 20~mA value for $\pm$2\% non-linearity given in the Hamamatsu datasheet. No deviations from the PMT response linearity are observed at the $\sim$20 nVs charge and $\sim$1 mA peak current scale of the CsI[Na] response to a 59.5 keV gamma ray from $^{241}$Am.

\begin{figure}[htbp]
\centering
\includegraphics[width=1.0\textwidth]{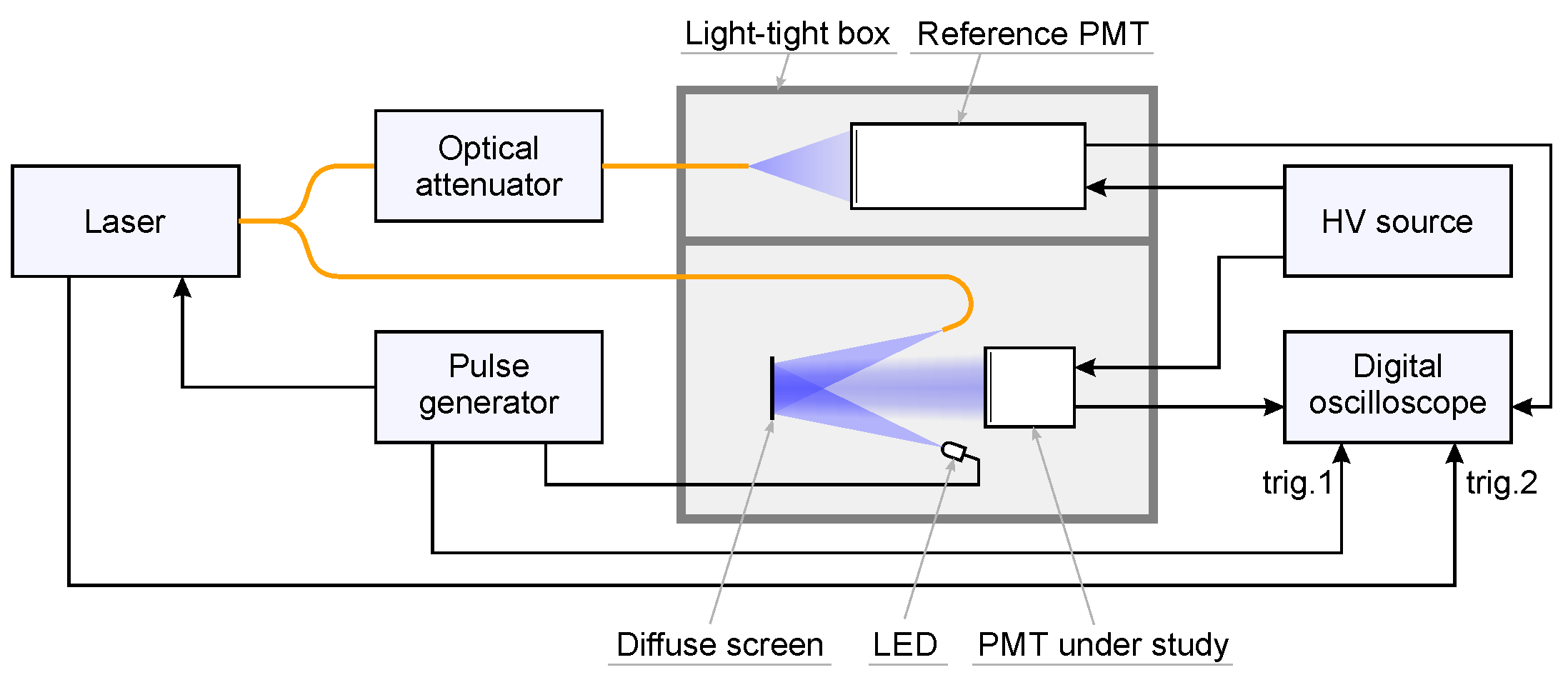}
\caption{\label{fig:Setup1} The experimental setup for study of the  H11934-200 PMT response in a two-pulse linearity test.}
\end{figure}
 
We also performed a two-pulse test to further probe the pulse linearity at a given energy scale. The basic idea of such a test is to measure the PMT response to light pulses of two kinds: a large pulse creating the instantaneous anode current of a certain scale (load) and a small pulse used to test the linearity (probe). One can measure characteristics of the PMT response to pulses of each kind first separately and then at the same time --- on top of each other. A mockup formula for a suitable linearity parameter could be

\begin{equation}
  \label{eq:two_pulse_1}
  R =  \frac{(Load \oplus Probe)-Load-Probe}{Probe},
\end{equation}
where ``$Load$'' and ``$Probe$'' correspond to an amplitude or a charge of the PMT response to respective light pulses, and $\oplus$ means that pulses appear on the top of each other. We used a 750~ns long LED pulse inducing a PMT signal of about 1400~PE as a load. Such a signal has an amplitude larger or similar to that induced by the 59.5 keV calibration line in our QF measurements and a larger total charge. A 30~ns long laser pulse with an integral of about 10 PE was used as a probe. Both LED and laser light illuminating the PMT under study was reflected from a diffuse screen for the best light uniformity. Stability of the 405~nm laser was controlled by the reference PMT (model FEU-143). The repetition rate of both laser and LED pulses was set to 100~Hz. The scheme of the experimental can be found in Figure~\ref{fig:Setup1}.

\begin{figure}[ht]
\centering
\includegraphics[width=1.0\textwidth]{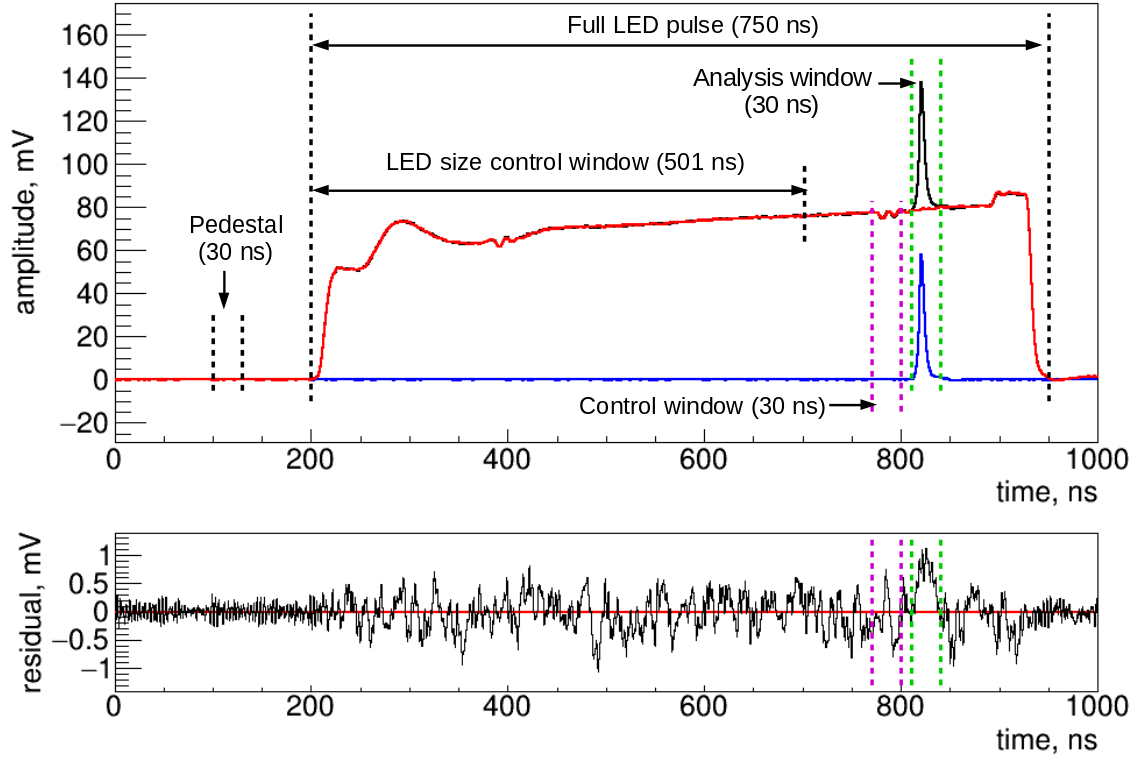}
\caption{\label{fig:two_pulse} An example of averaged waveforms obtained in the two pulse measurements (\#3 at -950V, see Table \ref{Table_TwoPulse}) and the residual amplitude plot. The signal was inverted and the pedestals were subtracted for clarity. Top: solid blue line - averaged laser signal ($Probe$), solid black line - averaged LED + laser signal ($Load \oplus Probe$), solid red line - averaged LED signal ($Load$); dashed lines mark signal integration windows. The averaged waveforms are corrected for the change in the LED charge (by the LED size control window) and the laser charge (by the reference PMT). The corrections are smaller than 1\% of the respective integrals. Bottom: the residual amplitude of the signals in the top panel, $(Load \oplus Probe) - Load - Probe$; dashed lines show the analysis (green) and control (magenta) time intervals; the red line marks 0 mV residual.}
\end{figure}

We performed five measurements of the $R$-parameter at each of four PMT bias voltage values: -935, -950, -980 and -1000~V. The analysis was based on integration of the averaged signal waveforms produced by the digital oscilloscope. Each averaged waveform represents either $\sim$5000 ($Probe$) or $\sim$10000 ($Load$, $Load \oplus Probe$) individual signal waveforms. We used short, 30~ns integration windows to obtain $R$. That allows us to suppress influence of large fluctuations of the full LED signal charge on the residual of interest. An example of the averaged waveforms used to measure $R$ is shown in Figure~\ref{fig:two_pulse} together with the definition of charge evaluation windows. It presents an interval used to correct for possible change in the LED intensity between $Load$ and $Load \oplus Probe$ measurements and a shifted 30~ns control window used as a sanity check for $R$ in the absence of a laser pulse. The results of the measurements are summarized in Table~\ref{Table_TwoPulse}.

\begin{table}{hbp}
\begin{center}
\parbox{0.85\linewidth}{\caption{\label{Table_TwoPulse} Results of the two pulse method measurement of the linerity parameter R at different PMT bias voltages (BV)}}
\begin{tabular}{ccccc}
\hline
BV, V & \multicolumn{2}{c}{Analysis window} & \multicolumn{2}{c}{Control window} \\
& R, \% & Mean $\pm$ RMS, \% & R, \% & Mean $\pm$ RMS, \% \\
\hline
-935  & 3.9, 5.1, 5.0, 5.0, 2.9 & 4.4 $\pm$ 1.0 &
2.1, 0.0, 0.1, 1.4, -0.2 & 0.7 $\pm$ 1.0 \\
-950  & 5.2, 4.4, 4.0, 3.0, 3.1 & 3.9 $\pm$ 1.0 &
3.3, -0.3, -1.0, 0.2, 0.3 & 0.5 $\pm$ 1.6 \\
-980  & 4.5, 2.1, 0.4, 4.8, 1.2 & 2.6 $\pm$ 2.0 &
-1.1, 0.6, -0.7, 0.1, -2.3 &  -0.7 $\pm$ 1.1 \\
-1000 & 5.6, 3.3, 3.4, -0.3, 2.2  & 2.8 $\pm$ 2.1  &
1.2, 1.8, 0.9, 1.8, 0.4 & 1.2 $\pm$ 0.6
\end{tabular}
\end{center}
\end{table}

It can be seen that the $R$ values evaluated for the analysis window are close to 3-4\% in average, while the average of values from the control window is close to 0\%. The laser probe appears to be slightly larger on the top of the LED signal than otherwise. This effect is smaller and of the opposite sign relative to the claim of ref.~\cite{Collar_2019}. The estimate of the $R$ uncertainty expected from the fluctuations in number of PE in a 30 ns window is 1.3\%. This estimate is close to the observed RMS of the measured values. The results suggest that the deviation is significant and can be interpreted as an overlinearity, although it is hard to make a conclusion about its dependence on the scale of a signal or the bias voltage of the PMT. It might be that the effect manifests itself only after the significant charge is generated by the PMT preceding the probe and its influence on the full signal integral is smaller than 3\%. That would explain why the results of the test with the reference PMT gave larger pulse linearity limits than the scale of the signals from the two-pulse test. Another explanation can be that the generator pulse powering the LED induced distortion in the bias voltage of the H11934-200 PMT or its first dynode potential. Indeed, during the course of the measurement the LED was placed side by side with the H11934-200 (see Figure~\ref{fig:Setup1}) and the voltage pulse amplitude was about +1.1~V. The distortion required to produce a $\sim$1\% effect in the vicinity of -950 V is 1~V in accord with the gain curve of the PMT; larger deviations may be caused if distortion affected potential at the first dynode. Despite some ambiguity in interpretation of results, the $R$-values measured in the two-pulse test contradict the $\sim15\%$ saturation reported in ref. \cite{Collar_2019}.

\begin{figure}[htbp]
\centering
\includegraphics[width=0.6\textwidth]{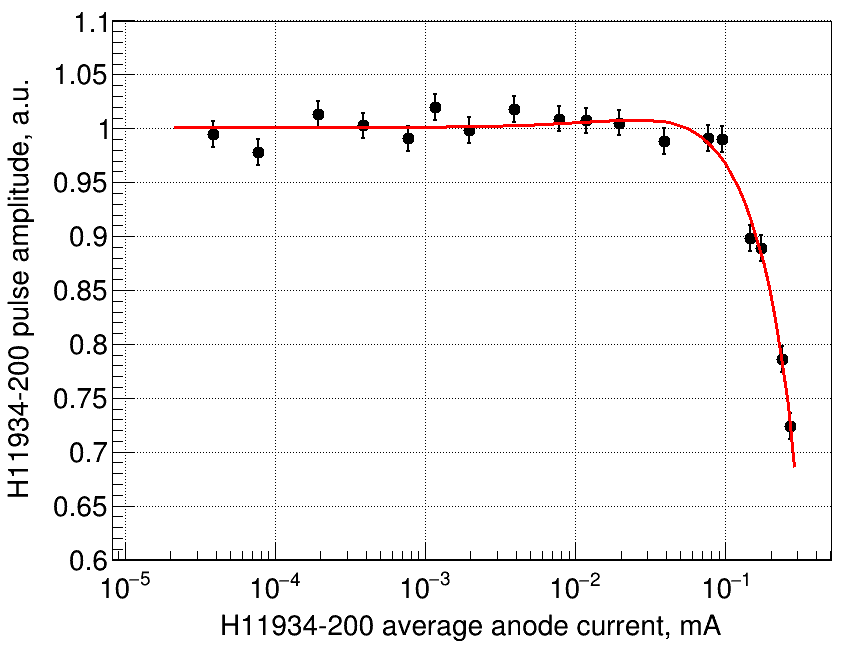}
\caption{\label{fig:nl_cur} Test pulse amplitude in the H11934-200 vs. the average anode current. The red line shows a fit to an ad hoc fourth degree polynomial. The uncertainty estimates were evaluated as the RMS of the data in the 10~nA - 0.1~mA range.}
\end{figure}

Finally, we measured the H11934-200 load capacity in terms of the average anode current according to the scheme in Figure~\ref{fig:Setup1}. The measurement was performed at -1000~V bias voltage. The 1.5~$\mu$s-long LED pulse illuminated the H11934-200 PMT at different repetition rates, increasing over the course of the measurement. The amplitude of a single LED pulse was about 50~mV in the H11934-200 unit. For each LED repetition rate, a short laser pulse illuminated both H11934-200 and the reference PMT at a fixed rate of 100~Hz. For each new measurement, the laser pulse was delayed by a given time to be sure no LED pulse immediately preceded the laser. This allowed us to measure and compare the amplitudes (and thus charge) of the laser signals in the H11934-200 and the reference PMT as a function of the LED repetition rate. The repetition rate was then converted to average anode current by multiplying the charge of a single LED pulse and the repetition rate with negligible contribution of the laser pulses. The results of the measurement are shown in Figure~\ref{fig:nl_cur}. Non-linearity manifests at a scale of $\sim$~0.08~mA of average anode current which is of the same order of magnitude as the 0.018~mA maximum rating of average anode current from the datasheet~\cite{H11934}. None of our measurements of CsI[Na] quenching factors or linearity tests with CsI[Na] crystal reached near such a value and thus are in the linear range for average anode current.

\subsection*{Summary of the PMT linearity tests} \label{subsec:lin_summary}

\begin{table}[bh]
    \caption{\label{Table_Ln_Sum} Summary table of the H11934-200 linearity tests.}
    \begin{tabular}{  p{3.0cm} p{2.2cm} p{8.5cm} }
\hline
\centering Test & \centering Reference & Result \\ \hline
\centering CsI[Na] relative light yield 
& \centering Figure~\ref{fig:rel_ly}
& Consistent with the published data up to 662 keV (about 7000~PE) within the uncertainties. An indication of the light yield about 3\% lower than the best-fit models from ref.~\cite{Beck_2015, Salakhutdinov_2015} is observed for 160-477 keV. No dependence of the effect on the PMT bias voltage.
\\ \hline
Pulse linearity by the reference PMT
&  \centering Figure~\ref{fig:nl_pulse}
& Saturation of about 3\% at 10~mA anode current --- about 400 nVs or 10000~PE of a 750 ns long signal at -950~V PMT bias voltage.
\\ \hline
Two pulse method
& \centering Table~\ref{Table_TwoPulse}, Figure~\ref{fig:two_pulse}
& An overlinearity-like effect of 3-4\% is observed in the late part of a $\sim$1400~PE 750 ns long signal. It is not clear if the size of the effect depends on the probe pulse position or the PMT bias voltage value.
\\ \hline
\centering Average anode current
& \centering Figure~\ref{fig:nl_cur}
& Saturation of about 2\% at the average anode current of 0.08~mA at -1000~V bias voltage.
\\ \hline
    \end{tabular}
\end{table}

We performed linearity tests of the H11934-200~PMT unit used for the COHERENT CsI[Na] quenching factor measurements (see the summary in Table~\ref{Table_Ln_Sum}). Our results suggest that the PMT signals induced by 59.5~keV gamma rays in the crystal are within the charge linearity region of the PMT. The pulse linearity limit is observed at the PMT response scale about 10 times higher than that, which is close to the estimate from the manufacturer. We refute the PMT non-linearity claim from ref.~\cite{Collar_2019} and disagree with corresponding corrections applied to our data. A conservative estimate of the linearity measurements accuracy which can be propagated to the QF uncertainties in quadrature is~$\pm3\%$.  Such a contribution is suggested by the results of the relative light yield and the two pulse tests. It would increase all of the QF uncertainties in this work by about 1\%. Such an increase would have negligible effect on the results of the global QF fit from Section~\ref{sec:global_fit} and would cause a fraction of percent change in the prediction of CEvNS rate in the COH-CsI detector. We do not propagate this uncertainty to Tables \ref{Table_COH1}, \ref{Table_COH2}, \ref{Table_COH3} or Figure\ref{fig:QFfit}.

\subsection{Light yield estimates and SPE charge distribution model} \label{subsec:LY_and_SPE}

\begin{figure}[htbp]
\centering
\includegraphics[width=1.0\textwidth]{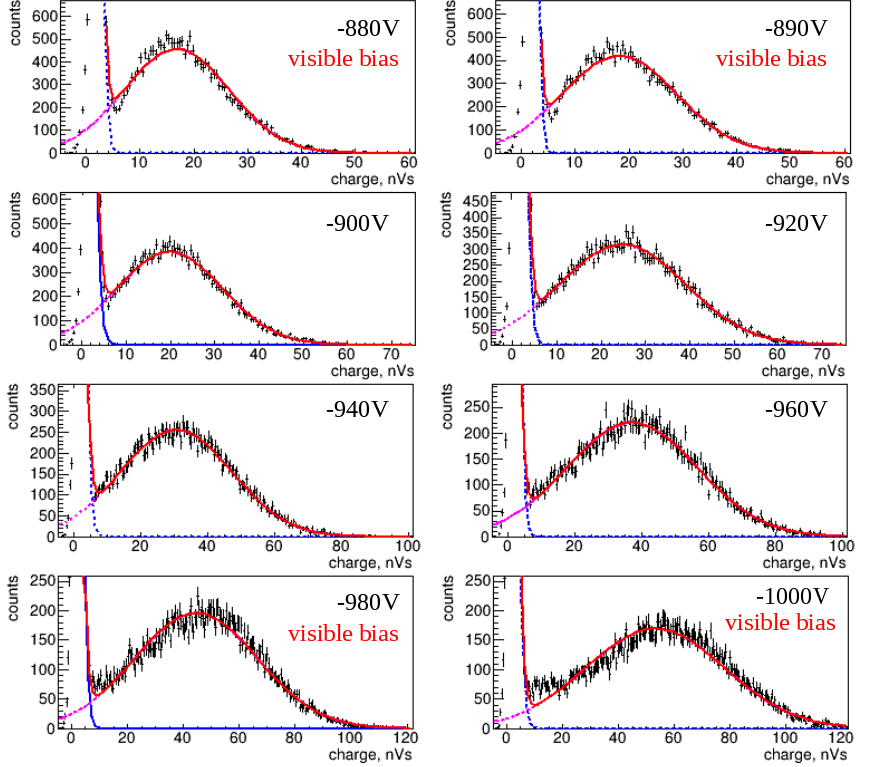}
\caption{\label{fig:spe_nl} Illustration of SPE charge spectra fits using a Gaussian-based SPE model with exponential background. Dotted blue and magenta lines correspond to exponential and Gaussian parts of the fit respectively.}
\end{figure}

\begin{figure}[htbp]
\centering
\includegraphics[width=0.9\textwidth]{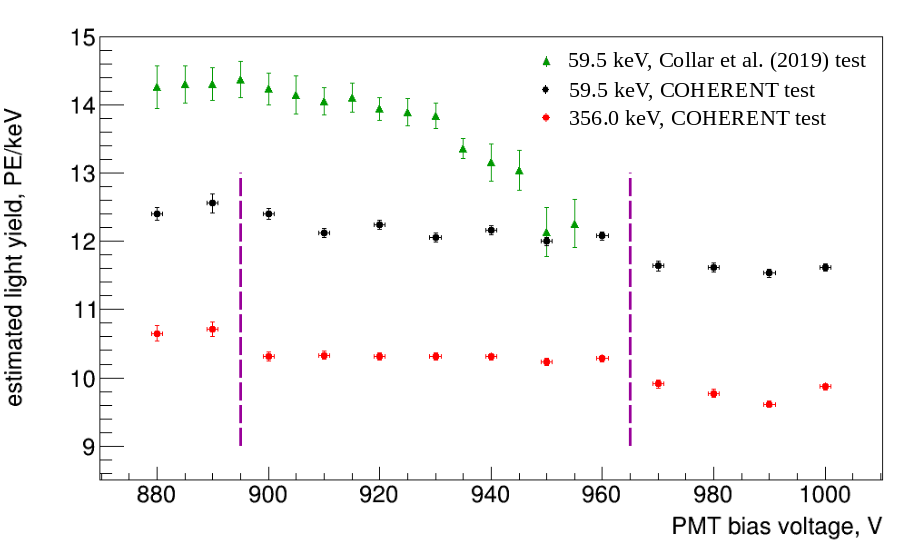}
\caption{\label{fig:abs_ly} The CsI[Na] absolute light yield estimates at different H11934-200 bias voltages for 59.5~keV~(black dots) and 356~keV~(red dots) in COHERENT tests. The green triangles show a similar plot for 59.5 keV redrawn from \cite{Collar_2019}. The estimates are evaluated with a Gaussian-based SPE charge model. Outside of the magenta lines, we fit only regions of the spectra in the vicinity of the SPE peak as the Gaussian-based model does not describe full spectra (COHERENT test).}
\end{figure}

The claim of the PMT non-linearity from ref.~\cite{Collar_2019} was based on the measurement of the CsI[Na] light yield estimate at 59.5 keV in units of PE/keV at different PMT bias voltage. We performed similar tests to see if the decreasing trend from ref.~\cite{Collar_2019} can be reproduced. Firstly, we measured the CsI[Na] light yield with $^{241}$Am and $^{133}$Ba sources while changing the PMT bias voltage in the range of -[880,1000]~V with 10~V steps. We used integrals from a 3~$\mu$s integration window of the 59.5~keV $^{241}$Am and 356~keV $^{133}$Ba lines. To determine the SPE charge, we used pulses which appear 60~$\mu$s after the start of a 59.5~keV signal. The PE density at this offset is low enough to neglect probability of multiple PE overlap. For each SPE pulse, we used a fixed length integration window from -2 ns to +8 ns relative to the maximum amplitude sample. This scheme allowed us to maintain consistency of integration approach for SPE signals of different amplitude scale, but similar shape, at different PMT bias voltages. We fit the SPE charge spectrum with a sum of an exponential distribution, representing electronic noise, and a Gaussian distribution for SPE charge. To the best of our knowledge a Gaussian-based approach was used in ref.~\cite{Collar_2019}. The results of such fits are presented in Figure \ref{fig:spe_nl}. One can see that the shape of the SPE charge distribution deviates significantly from the Gaussian below -900 V and above -970 V. While the former demonstrates significant asymmetry and may be affected by the pulse selection amplitude threshold in the low integral regime, the latter shows deviations from a Gaussian shape, especially in the valley between signal and noise. We also note that the Gaussian fits at all bias voltages demonstrate significant tails at negative integrals, which is unphysical. The fraction of the Gaussian integral in this tail varies from 4.2\% at -880 V to 1.5\% at -1000 V. In general such tails can originate from convolution of a signal-related distribution with the noise pedestal, but given the noise characteristics in the measurement discussed and the length of integration window (10 ns) this is not the case. We conclude that a pure Gaussian description of the SPE spectrum is not representative and can induce bias in the evaluation of absolute light yield, especially in combination with threshold effects of the pulse finding and integration approach (see Appendix~\ref{subsec:spe_int}). We use the results of the fits from Figure~\ref{fig:spe_nl} to obtain the absolute light yield estimates in units of PE/keV. For the cases of -880 V, -890 V, -970 V, -980 V, -990 V and -1000 V we restrict the fit range to the vicinity of the SPE peak to capture its position. The results are shown in Figure~\ref{fig:abs_ly}.

We observe decreased light yield estimates with increasing bias voltage, though the effect is much less than the $\sim13$\% for 900-950V bias voltage range suggested by ref. \cite{Collar_2019}. We demonstrate that the observed trend is not related to the size of a signal -- the light yield for a 356~keV signal behaves similarly to that for 59.5~keV. The same test with the 356~keV line addresses concerns that the lower light yield seen in the current tests is significant enough to push the 59.5~keV line out of the non-linear regime (a light yield $\sim$~12~PE/keV in this test vs. $\sim$~14~PE/keV for 59.5~keV in ref. \cite{Collar_2019}). Similar behavior for 59.5~keV and 356~keV lines light yield estimates suggests that the observed downwards trend is related to the evaluation of the SPE charge rather than the magnitude of the calibration line.

\begin{figure}[htbp]
\centering
\includegraphics[width=1.0\textwidth]{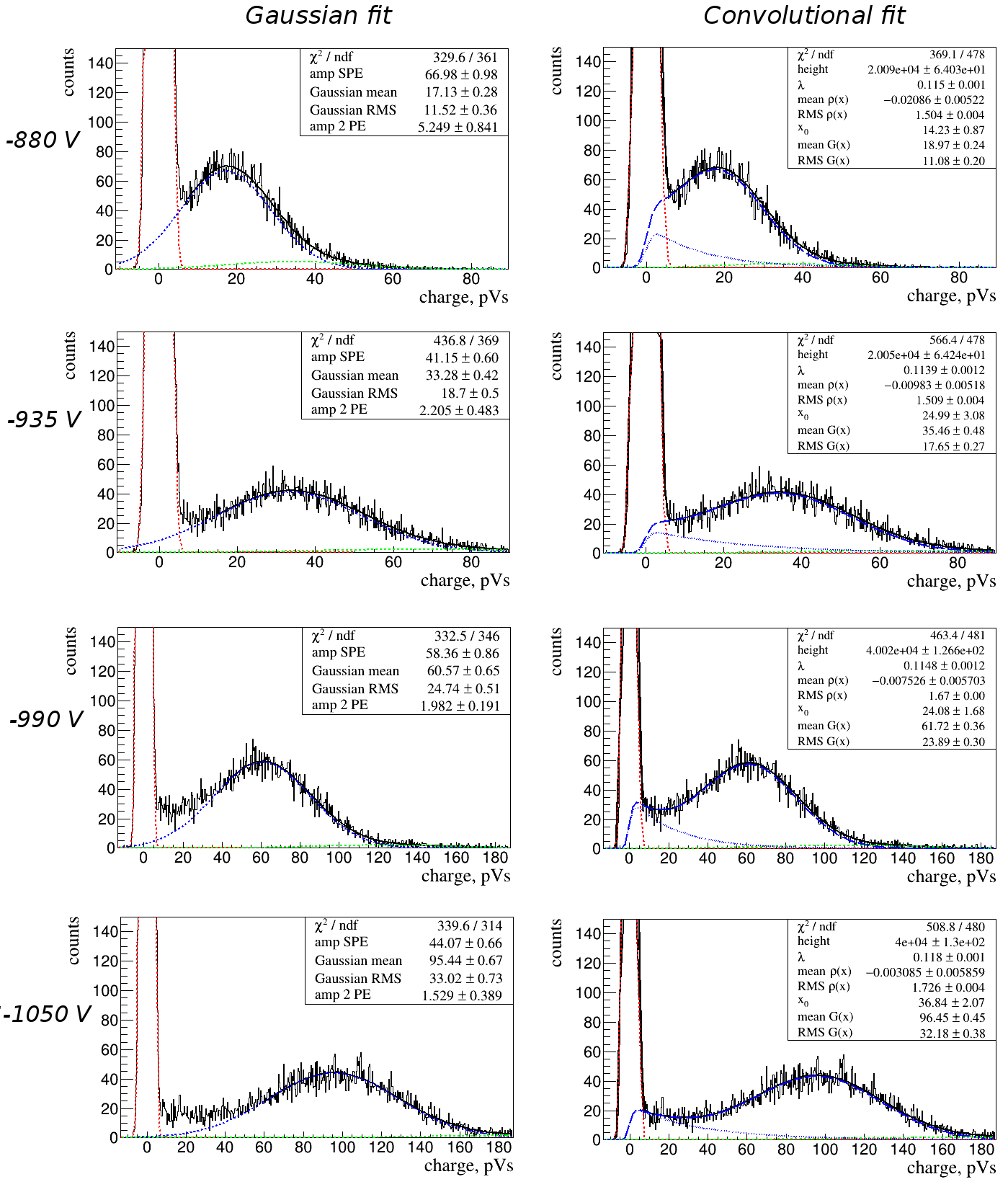}
\caption{\label{fig:nl_spe} Examples of SPE spectra fit to the Gaussian SPE model in the spectral peak range (left) and the convolutional model taking into account contribution of under-amplified PE signals (right)}.
\end{figure}

In order to reinforce our conclusions about the possible bias in the SPE charge evaluation approach, we perform additional PMT tests with controlled light sources allowing threshold-less integration of SPE pulses. The sources are a 470~nm LED and a 405~nm picosecond laser pulse generator. The PMT under study was illuminated with both LED and laser light, reflected from a diffuse screen for the best light uniformity. We also used a reference PMT (FEU-143 model). A scheme of the experimental set-up is similar to one shown in Figure~\ref{fig:Setup1}. We measured the PMT response to the light pulses of constant intensity and the SPE charge spectra in the bias voltage range from -880~V to -1050~V. The pulser was used to trigger 1.5~$\mu$s-long laser pulses to be distributed to both the PMT under study and the reference PMT through an optical fiber splitter. The reference PMT was used primarily to correct for laser light instabilities during data-taking. The PMT under study detected around~3400 PE from a laser light pulse at 100 Hz rate. To measure the SPE spectra, the LED was biased with short, $\sim$5~ns pulses for low-intensity illumination of the PMT. Note that these spectra were obtained with no-threshold integration of a 20~ns time window aligned with the start of the LED pulse, triggered by the pulser. By looking at the spectra obtained at higher bias voltages we conclude that the SPE charge distribution is non-Gaussian with a non-trivial population in the valley between signal and noise as shown in Figure~\ref{fig:nl_spe}. This quasi-flat valley region is light-related --- the charge spectra obtained without LED pulses do not suggest a pedestal contribution to this region. These are most likely signal pulses that are under-amplified.

\begin{figure}[htbp]
\centering
\includegraphics[width=0.7\textwidth]{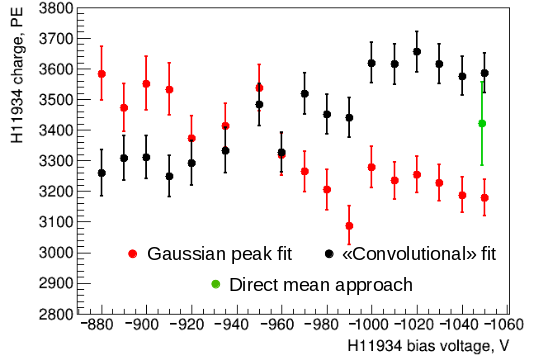}
\caption{\label{fig:laser_ly} The H11934-200 PMT response to the signal with a magnitude of $\sim3400$ PE vs. bias voltage. The units of PE are evaluated with different SPE fitting models (see text).}.
\end{figure}

Further we compare results of two SPE evaluation approaches. The first is Gaussian-based. We fit available spectra with the sum of a Gaussian distribution, taking into account double-PE pulses by adding a second Gaussian with twice the mean and twice the variance. The pedestal is fit separately with a Gaussian distribution to evaluate the baseline charge, set to 0~nVs in Figure~\ref{fig:nl_spe}. This model, even if not able to describe the whole spectrum range, is intended to represent the approach used in our studies with the CsI[Na] crystal and is close to the treatment in ref. \cite{Collar_2019}, to the best of our knowledge. The second model tested is adopted from ref. \cite{Alexander_2013} and is based on the sum of a pedestal Gaussian, an exponential, intended to represent under-amplified or under-collected PE, and a truncated Gaussian representing SPE charge. The latter two components are also convolved with the pedestal Gaussian to take into account the effect of the electronic noise on the SPE charge spectrum shape. The exact expression of the fit is

\begin{equation}
  \label{eq:spe1}
  F(x)=\sum_{n=0} P(n;\lambda)f_n(x),
\end{equation}

\begin{equation}
  \label{eq:spe2}
  f_n(x)=\rho(x)*\psi_1^{n*}(x),
\end{equation}

\begin{equation}
  \label{eq:spe3}
    \psi_1(x)= 
\begin{cases}
    p_E(\frac{1}{x_0}e^{-\frac{x}{x_0}})+(1-p_E)G(x;\mu,\sigma),& x>0\\
    0,              & x\leq 0
\end{cases}
\end{equation}
where $P(n;\lambda)$ is a Poisson distribution with a mean of $\lambda$, $\rho(x)$ gives the noise contribution (Gaussian pedestal), ``$*$" represents the convolution operator, ``$n*$" gives the distribution convolved with itself $n$ times and $G(x;\mu,\sigma)$ is a Gaussian with a mean $\mu$ and RMS of $\sigma$. We fix the $p_E$ parameter to be 0.2 based on fits at larger absolute bias voltage as the exponential contribution becomes indistinguishable from the pedestal at lower gains and consider the numerical mean of $\psi_1(x)$ to represent the mean SPE charge. Examples of the acquired SPE spectra are shown in Figure~\ref{fig:nl_spe} together with fits to $F(x)$ and a Gaussian fit of the SPE peak. The results of these fits were utilized to produce Figure~\ref{fig:laser_ly}.

It may be seen that results of Gaussian fits reproduce the trend from our studies with the CsI[Na] crystal (Figure \ref{fig:abs_ly}) showing a $\sim$10\% decrease from -880 V to -1050 V bias voltage. Gaussian fits at lower absolute bias voltage values have a non-negligible negative part biasing the mean SPE charge estimate to lower values, while at higher absolute bias voltage these fits do not describe the valley region and provide an overestimated mean SPE charge value. Combined, these effects create an image of decreasing trend and may be misinterpreted as evidence of non-linear behavior of the PMT, while being a consequence of incorrect SPE spectral shape model. Aside from the Gaussian model, there are other models in the literature including the SPE spectrum description with a Polya distribution \cite{Prescott_1966}, one or several gamma distributions \cite{DEAP_3000}, several Poisson distributions \cite{Degtiarenko_2017} or combination of a truncated Gaussian with an exponent \cite{Alexander_2013} described above. As can be seen in the case of our measurement the latter yields a trend opposite to the simple Gaussian model. By using this alternative approach we demonstrate that choice of SPE model can affect the mean SPE charge estimates on the order of $\sim$10\%, which was also pointed out in ref. \cite{Saldanha_2017,Anthony_2018}, and lead to bias in the absolute light yield estimates.

While a test of SPE models variety presented in the literature is beyond the scope of this work, we can make an estimate of the SPE mean based on the charge distribution at the largest bias voltage of -1050~V (see bottom panel in Figure~\ref{fig:nl_spe}). The mean of this distribution in the range from 7~pVs (negligible pedestal contribution) to 187~pVs (the upper edge of the histogram) can serve as a reasonable estimate of the SPE mean charge. The combined effects of events absorbed by the pedestal, a limited histogram range and a contribution from double PE pulses result in $\pm$4\% uncertainty on the SPE mean at -1050 V. The corresponding signal size in PE is shown in Figure~\ref{fig:laser_ly} in green. While the precision of this estimate is still limited it constrains the size of potential variations large enough to verify the obtained signal scale in units of PE.

\subsection*{Summary of a SPE model effect on light yield estimates} \label{subsec:LY_and_SPE_sum}

Following ref.~\cite{Collar_2019} we measured the PMT response to the CsI[Na] scintillation from 59.5 keV gamma rays in units of PE/keV at different bias voltages. We completed this measurement with a similar test for a 356.0 keV line and, separately, for the laser pulses of fixed intensity. We observe a decreasing trend of the absolute light yield estimates with increase of the PMT bias voltage, but prove that it does not depend on the size of a signal (Figure~\ref{fig:abs_ly}) as it would be in the case of the PMT non-linear behavior. We also show that the observed trend of absolute light yield as a function of PMT bias voltage significantly depends on the SPE charge distribution model utilized (Figure~\ref{fig:laser_ly}) and should not be used for the characterization of PMT response linearity. The Gaussian model resulting in the decreasing trend fails to describe experimental SPE charge distributions in the large bias voltage range considered and can bias the signal size estimates in units of PE (Figure~\ref{fig:spe_nl}, \ref{fig:nl_spe}). The absolute light yield estimates from Table~\ref{Table_LY} evaluated with a Gaussian SPE charge model may be affected by this ambiguity, while the measured QF values do not depend on the SPE charge model. The mean PE charge estimate, whichever model is used, cancels out in the definition of QF. We do not have an explanation for the dependence of the absolute light yield studied in ref. \cite{Collar_2019}, although our results suggest that bias related to the SPE charge-fitting model may contribute to it.

\end{document}